\documentclass[aps,prl,twocolumn,superscriptaddress,amsmath,amssymb]{revtex4}
\usepackage{graphicx}
\usepackage{dcolumn}
\usepackage{bm}
\usepackage{color}
\usepackage{amsmath}
\usepackage{bbold}

\usepackage{kantlipsum,setspace}
\usepackage[justification=raggedright,font=small,margin=0pt]{caption}

\renewcommand{\vec}[1]{{\boldsymbol #1}}
\begin{document}

\title{
\begin{center} 
\bf Planar Hall effect from the surface of topological insulators
\end{center} 
}

\author{A. A. Taskin}
\affiliation{Physics Institute II, University of Cologne, Z{\"u}lpicher Str. 77, 50937 K{\"o}ln, Germany}
\author{Henry F. Legg}
\affiliation{Institute for Theoretical Physics, University of Cologne, Z{\"u}lpicher Str. 77, 50937 K{\"o}ln, Germany}
\author{Fan Yang}
\affiliation{Physics Institute II, University of Cologne, Z{\"u}lpicher Str. 77, 50937 K{\"o}ln, Germany}
\author{Satoshi Sasaki}
\altaffiliation[Present address:]{School of Physics and Astronomy, University of Leeds, UK}
\affiliation{Institute of Scientific and Industrial Research, Osaka University, Mihogaoka 8-1, Ibaraki, 567-0047  Osaka, Japan}
\author{Yasushi~Kanai}
\affiliation{Institute of Scientific and Industrial Research, Osaka University,  Mihogaoka 8-1, Ibaraki, 567-0047  Osaka, Japan}
\author{Kazuhiko Matsumoto}
\affiliation{Institute of Scientific and Industrial Research, Osaka University, Mihogaoka 8-1, Ibaraki, 567-0047  Osaka, Japan}
\author{Achim Rosch}
\affiliation{Institute for Theoretical Physics, University of Cologne, Z{\"u}lpicher Str. 77, 50937 K{\"o}ln, Germany}
\author{Yoichi Ando}
\affiliation{Physics Institute II, University of Cologne, Z{\"u}lpicher Str. 77, 50937 K{\"o}ln, Germany}

\renewcommand{\abstractname}

\begin{abstract}
{ \bf A prominent feature of topological insulators (TIs) is the surface states comprising of spin-nondegenerate massless Dirac fermions \cite{Hasan-Kane, Qi-Zhang, Ando}. Recent technical advances have made it possible to address the surface transport properties of TI thin films while tuning the Fermi levels of both top and bottom surfaces across the Dirac point by electrostatic gating \cite{Dual-gate}. This opened the window for studying the spin-nondegenerate Dirac physics peculiar to TIs. Here we report our discovery of a novel planar Hall effect (PHE) from the TI surface, which results from a hitherto-unknown resistivity anisotropy induced by an in-plane magnetic field \cite{AMR_3d}. This effect is observed in dual-gated devices of bulk-insulating Bi$_{2-x}$Sb$_{x}$Te$_{3}$ thin films, in which both top and bottom surfaces are gated. The origin of PHE is the peculiar time-reversal-breaking effect of an in-plane magnetic field, which anisotropically lifts the protection of surface Dirac fermions from back-scattering. The key signature of the field-induced anisotropy is a strong dependence on the gate voltage with a characteristic two-peak structure near the Dirac point which is explained theoretically using a self-consistent T-matrix approximation. The observed PHE provides a new tool to analyze and manipulate the topological protection of the TI surface in future experiments.}

\end{abstract}

\maketitle

The two-dimensional (2D) Dirac fermions on the surface of TIs are immune to localization by a random scalar potential \cite{Nomura} and are often said to be ``topologically protected". Besides the bulk-edge correspondence of a topological system to guarantee the gapless nature \cite{Ando}, there are two closely related reasons for this protection: 
First, the $\pi$ Berry phase associated with massless Dirac fermions protects them from weak localization effect \cite{TAndo}. Second, the spin is perpendicularly locked to the momentum, which suppresses the back-scattering on non-magnetic scatterers \cite{Hasan-Kane, Qi-Zhang, Ando}. However, the topological protection can be lifted in several situations. For example, when a sample is too thin and the wavefunctions of the top and bottom surface states overlap, the hybridization between the two opens up a gap at the Dirac point \cite{QiKunXue}, leading to a loss of topological protection \cite{Taskin_MBE}. Also, since time-reversal symmetry (TRS) is the prerequisite of topological states in TIs \cite{Hasan-Kane, Qi-Zhang, Ando}, breaking of TRS is another way to lift the topological protection. Applying a magnetic field perpendicular to the TI surface introduces a mass term in the Dirac Hamiltonian to open up a gap in the surface states. When the field is applied along the surface of a TI, TRS is also broken, but such a parallel magnetic field will not affect helical surface states besides a shift of the Dirac dispersion in the momentum space; hence, no gap will open in the Dirac dispersion for high-symmetry orientations of the surface and the magnetic field (see Supplementary Information \cite{SI} for more details.)

Our experiment is designed to address the effect of the parallel magnetic field on the surface transport, when TRS is broken but the massless Dirac state is preserved. We found that the scattering of Dirac fermions in this situation becomes anisotropic, because the spin-momentum locking causes a difference in the scattering amplitudes for particles with the spin parallel and perpendicular to the magnetic-field direction. This leads to a magnetic-field-induced anisotropy in the resistivity measured along and perpendicular to the field, which results in a novel PHE. In other words, this intriguing effect is a manifestation of the momentum-selective lifting of the topological protection due to TRS breaking. 

\begin{figure}
\begin{center}
\includegraphics[width=8.7cm]{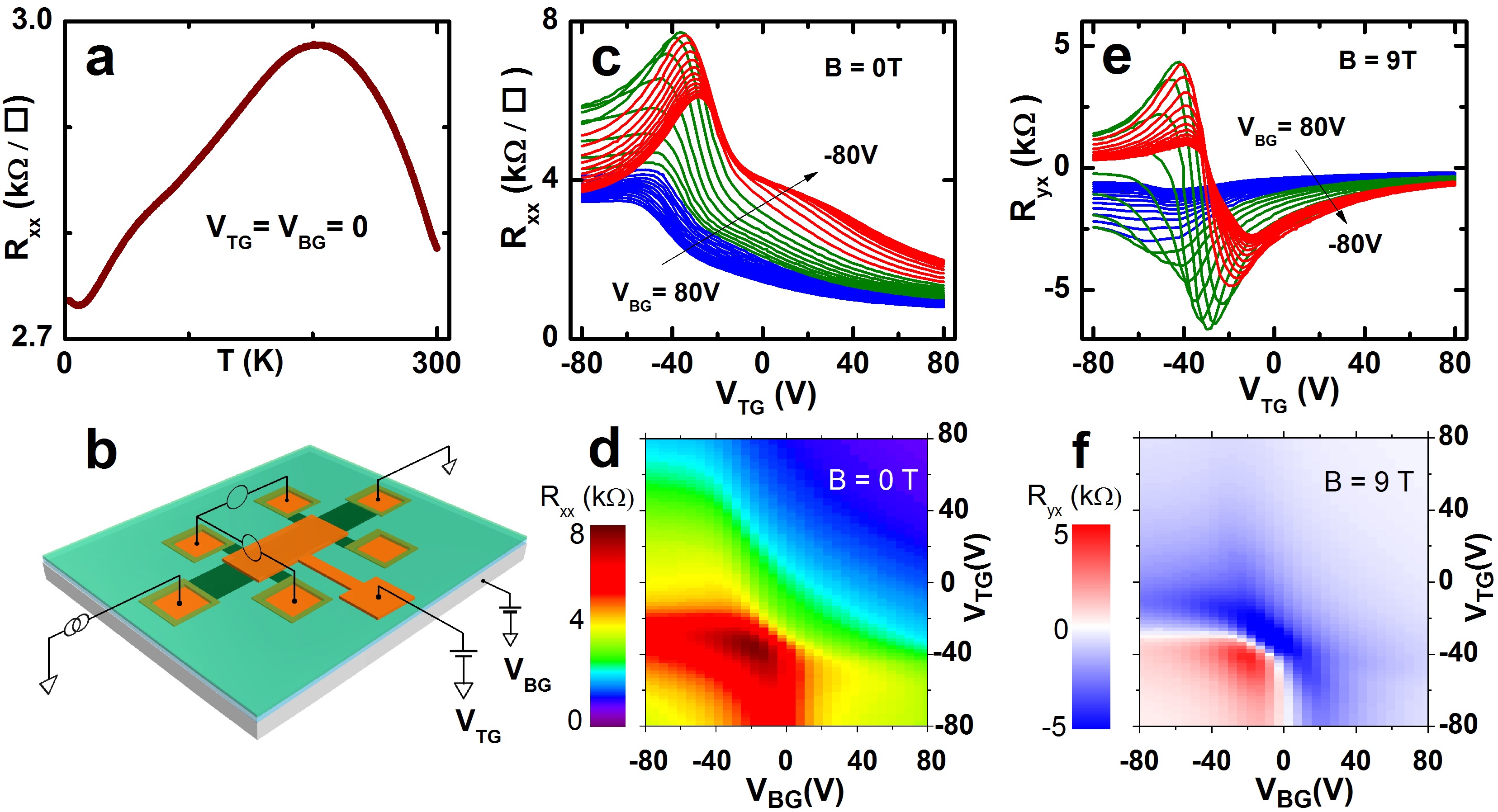}
\caption{
{\bf Dual-gating of BST films.} (a) Temperature dependence of $R_{xx}$ in a 17-nm-thick device at zero gate voltages ($V_{TG}$ = $V_{BG}$ = 0). (b) Schematics of the dual-gate Hall-bar device and the measurement configuration. (c), (d) Gate-voltage dependencies of $R_{xx}$ in 0 T at 2 K. (e), (f) Gate-voltage dependencies of $R_{yx}$ in the perpendicular magnetic field of 9 T at 2 K. 
} 
\end{center}
\end{figure}

To access the surface transport properties, one needs to suppress the bulk contribution in the total conductance. There are several ways to achieve this \cite{Ren_BTS, Cava_BTS, Fuhrer_BS, Ren_BSTS, Chen_BSTS} : The most effective one is the compensation of donors and acceptors in the TI material to bring the Fermi level into the bulk band gap. Reducing the thickness of a sample can also be  effective, due to a reduced bulk/surface ratio. At present, thin-film samples of TIs grown by the molecular beam epitaxy (MBE) technique are among the best for surface transport experiments \cite{AQHE, Oh_BS, Samarth_GAMR, Tokura_s-to-c}. For example, Bi$_{2-x}$Sb$_{x}$Te$_{3}$ (BST) thin films, in which the optimization of the composition can give almost perfect compensation, were used for studying the integer quantum Hall effect on the TI surface \cite{Tokura_BST}. For the present experiments, BST films with a bulk-insulating composition ($x \approx$ 1.7) were grown on sapphire by MBE. A typical temperature dependence of the sheet resistance $R_{xx}$ in a bulk-insulating sample is shown in Fig. 1a. Below about 200 K, the resistivity is dominated by metallic surface transport. The magnitude of $R_{xx}$ depends on the charge carrier density $n_s$ on the top and bottom surfaces of the film, which can be controlled by electrostatic gating \cite{Dual-gate}. Our dual-gate device, shown schematically in Fig. 1b, provides the ability to tune $n_s$ on both surfaces independently. For example, as shown in Fig. 1c, $R_{xx}$ can reach a high value of $\sim$8 k$\Omega$ by suitably tuning the top- and bottom-gate voltages, $V_{TG}$ and $V_{BG}$, respectively. The effect of this dual-gating can be clearly seen in the colour mapping shown in Fig. 1d, where the maximum in $R_{xx}(V_{TG}, V_{BG})$, corresponding to the dark-red region, signifies the simultaneous crossing of the Dirac points on both top and bottom surfaces. 
The Hall resistance $R_{yx}$ was measured in magnetic fields perpendicular to the films, and its gate-voltage dependencies are shown in Fig. 1e for $B$ = 9 T; here, one can see a sharp change between $n$- and $p$-type carriers in a specific range of gate voltages. The zero-crossing of $R_{yx}$, which can be easily recognized in the colour mapping shown in Fig. 1f as a white band separating red (p-type) and blue (n-type) regions, can be used as an indicator of the Dirac-point crossing of the Fermi level. 

Our main result, observation of the PHE, is shown in Fig. 2. For these measurements, the magnetic field was applied parallel to the film and was rotated within the film plane. The angle $\varphi$ between the field and the current direction is defined in the central inset of Fig. 2. The planar Hall resistance $R_{yx}$, i.e. the transverse resistance measured across the width of the sample perpendicular to the current [as shown in the inset of Fig. 2a], shows a non-zero value for all field directions except for the parallel and perpendicular orientations. In fact, it follows the $\sim$cos$\varphi\,$sin$\varphi$ angular dependence as exemplified in Fig. 2a for $V_{TG}$ = $V_{BG}$ = 5 V and $B$ = 9 T. Such a behaviour is not expected for non-magnetic materials. The 180$^{\circ}$-periodic angular dependence was also observed in the longitudinal resistance $R_{xx}$ as shown in Fig. 2b; this kind of resistivity oscillation is generally called anisotropic magnetoresistance (AMR). The observed AMR follows the $\sim$cos$^{2}\varphi$ angular dependence. Phenomenologically, both PHE and AMR stem from an anisotropy in the resistance tensor, and the observed $\varphi$ dependence is expected when the magnetic field sets the anisotropy axis, along which the resistance becomes larger (see Supplementary Information for details \cite{SI}). 

\begin{figure}
\begin{center}
\includegraphics[width=8.7cm]{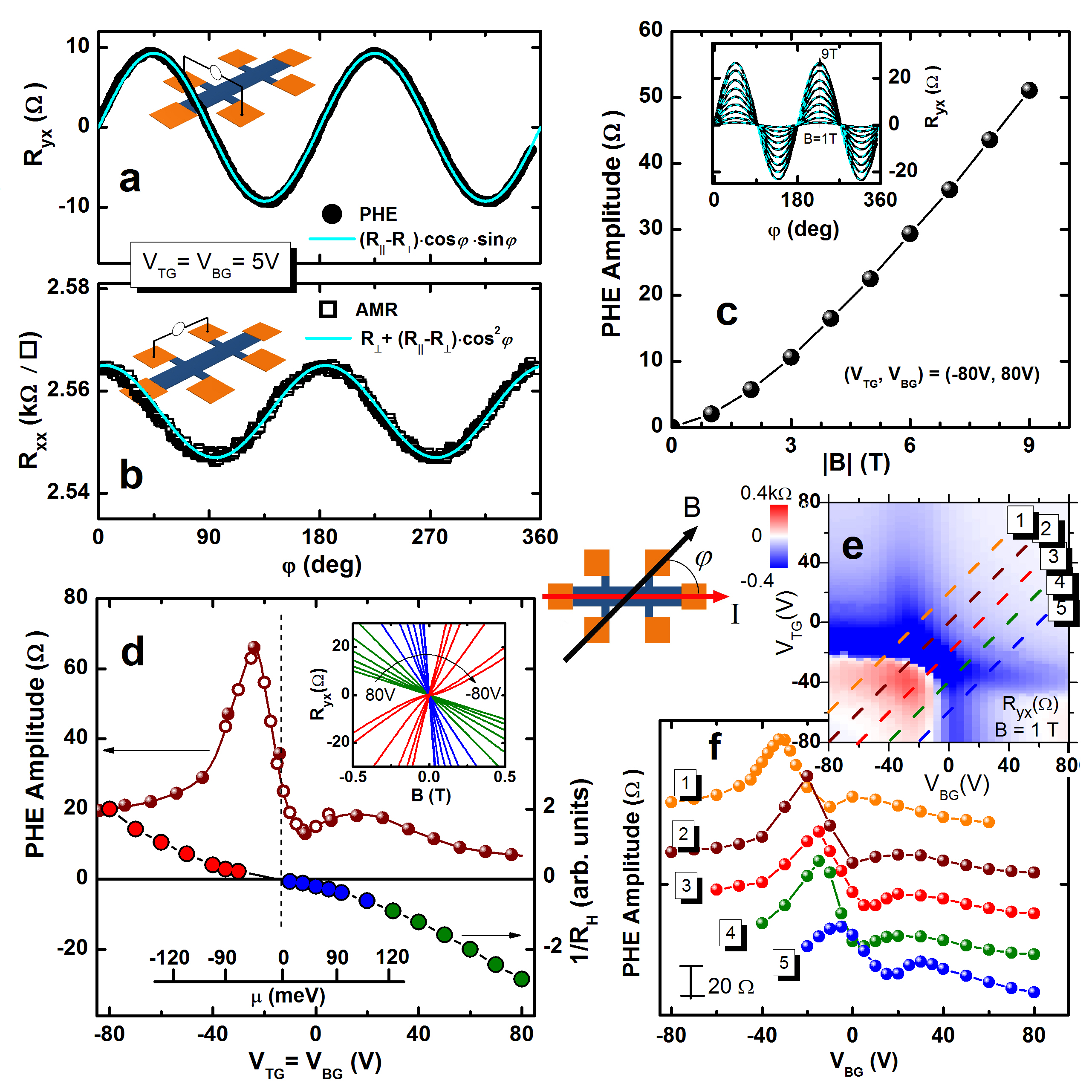}
\caption{
{\bf Planar Hall Effect}. (a) Angular dependence of planar $R_{yx}$ data at $V_{TG}$ = $V_{BG}$ = 5~V measured at 2~K in the magnetic field of 9 T rotated in the film plane (inset shows the configuration); blue solid line is a fit to ($R_{\parallel} - R_{\perp}$)cos$\varphi\,$sin$\varphi$, where $\varphi$ is defined in the central inset. (b) Angular dependence of $R_{xx}$ in the same conditions as in (a); blue solid line is a fit to ($R_{\parallel} - R_{\perp}$)cos$^{2}\varphi$. (c) Magnetic-field dependence of the PHE amplitude ($\equiv R_{\parallel} - R_{\perp}$) at $V_{TG}$ = $-80$ V and $V_{BG}$ = 80 V; inset shows the raw $R_{yx}(\varphi)$ data and their fits in various $B$. (d) Gate-voltage dependence of the PHE amplitude for $V_{TG}$ = $V_{BG}$ in the in-plane 9-T field (left axis) and the effective total carrier density (right axis), obtained from the low-field Hall data (shown in the inset); vertical dashed line marks the Dirac-point crossing, and the scale-bar inset depicts the estimated change of the Fermi level. (e) Colour mapping of $R_{yx}(V_{TG}, V_{BG})$ measured in the out-of-plane 1-T field, on which different dual-gating paths for the PHE-amplitudes measured in the in-plane 9-T field shown in (f) are indicated; curves in (f) are shifted for clarity.
} 
\end{center}
\end{figure}

According to the resistance-tensor phenomenology \cite{SI}, the amplitudes of PHE and AMR should be the same and are both written as $R_{\parallel} - R_{\perp}$, where $R_{\parallel}$ ($R_{\perp}$) is the sheet resistance for $B \parallel I$ ($B \perp I$). Nevertheless, due to a possible misalignment of the experimental plane of rotation with respect to the film plane, the observed AMR can be contaminated by the contribution from the ordinary orbital magnetoresistance $\Delta R^{*}_{\perp}$, which comes from a finite magnetic-field component $B^{*}_{\perp}$ perpendicular to the film. Although $B^{*}_{\perp}$ is normally very small, a large surface sheet resistance (up to several k$\Omega$) can cause the contribution from the orbital MR to become comparable to the amplitude of the AMR ($\sim$ several tens of $\Omega$). To make matters worse, upon the magnetic-field rotation, $B^{*}_{\perp}$ will change as $\sim \cos \varphi$ \cite{SI}, which, combined with the $\Delta R^{*}_{\perp} \sim B^{2}$ behaviour expected for the orbital MR, causes the spurious signal to present a $\sim$cos$^{2}\varphi$ dependence; this is virtually indistinguishable from the genuine AMR signal, and hence the amplitudes of the AMR in actual experiments are not always reliable. On the other hand, the ordinary Hall contribution due to $B^{*}_{\perp}$ is antisymmetric with respect to $B$ and can be easily removed from the PHE signal by taking the data in both positive and negative $B$ \cite{SI}. Therefore, PHE gives the genuine amplitude of $R_{\parallel} - R_{\perp}$, and all quantitative discussions in this paper are based on the measurements of PHE. Figure 2c shows an example of the magnetic-field dependence of $R_{\parallel} - R_{\perp}$, which is super-linear up to 9 T and shows no sign of saturation.

In ferromagnets, the phenomenon of AMR has been known for a long time \cite{AMR_3d}; in fact, Lord Kelvin reported the AMR almost 200 years ago. The understanding of its origin was eventually established through works of Mott (1936) \cite{Mott}, Smit (1951) \cite{Smit}, Campbell, Fert, and Jaoul (1970) \cite{CFJ}, and extended by others. This effect is best understood in diluted magnetic alloys, where the coexistence of $s$- and $d$-bands near the Fermi energy and a strong spin-orbit coupling are the two main ingredients for the AMR. The rotation of the magnetization by the magnetic field changes the population of unoccupied $d$-states with respect to the current direction, leading to a change in the scattering rate between $s$- and $d$-bands. 
Clearly, this mechanism is not applicable to a non-magnetic TI investigated here. 

To address the origin of the PHE and AMR in TI films, we took advantage of our dual-gate capability to tune the density and the type of carriers (and hence their helicity) on both surfaces independently, to see how these parameters influence the observed anisotropy. It turns out that for both $n$- and $p$-type states, the anisotropy is always positive, i.e. $R_{\parallel} > R_{\perp}$. Moreover, when the Fermi level is moved through the Dirac point and the surface conduction is changed from $n$- to $p$-type, the PHE amplitude was found to present an unusual two-peak structure with a local minimum at the Fermi-level position close to the Dirac point. Figure 2d shows an example of the PHE amplitude vs gate voltage along the dual-gating path with $V_{TG}$ = $V_{BG}$, presenting two peaks and a minimum; this minimum is located near the gate voltage where the effective total carrier density (deduced from the low-field Hall coefficient $R_H$) becomes zero, which roughly corresponds to the Dirac-point crossing. Measurements along different dual-gating paths near the Dirac point indicated in Fig. 2e found essentially the same behaviour, apart from a slight broadening and a shift along the $V_{BG}$-axis related to the shift in the transition from the $n$- to $p$-type region; this means that the characteristic two-peak structure is always associated with the Dirac-point crossing. 

This result is in stark contrast to the result of the AMR measurements in exfoliated flakes of another compensated TI material, BiSbTeSe$_{2}$ \cite{Sulaev_Nano}, where the AMR amplitude was observed to change from positive to negative upon applying a gate voltage to a 160-nm-thick flake from a bottom gate and measuring $R_{xx}$ on the top surface. In this regard, we were able to reproduce similar behaviour in our devices by intentionally setting a misalignment angle of $\sim$1$^{\circ}$ upon rotation. In our series of control experiments, including simultaneous measurements of AMR and PHE and adoptations of different mounting configurations \cite{SI}, we found that the negative AMR is an artifact due to the orbital MR and is strongly gate-voltage dependent. This conclusion is also supported by the temperature dependences of AMR and PHE: their amplitudes merge at high temperature where the orbital MR amplitude diminishes (see SI for details). We note that the bulk contribution could also play a role in the negative AMR in a thick TI flake, because the longitudinal magnetoresistance can become negative in the bulk transport, as recently reported for Bi$_{2}$Se$_{3}$ films \cite{NLMR}.

The occurrence and sign of the observed AMR follows directly from the spin-momentum locking of Dirac fermions on the TI surface and the associated topological protection from backscattering in the absence of broken TRS, which can be lifted by an in-plane magnetic field. The backscattering is highly sensitive to the relative orientation of this field and the electron velocity, as is shown in the following theoretical calculations. 
 
When describing the AMR theoretically, one first notes that an in-plane and uniform magnetic field has no effect on the electrons if one  models the surface based on the 2d Dirac equation and potential scattering from disorder:  orbital effects  are absent and the Zeeman coupling can be gauged away by a simple shift of the Dirac point\cite{Burkov_in-plane, Yakovenko_in-plane}. In reality, however, the uniform magnetic field in a disordered medium with spin-orbit interactions will generate magnetic fields at random positions that can induce spin-flip scattering. To describe this effect we consider a two-dimensional model where the Dirac electrons
hybridize with impurities located at random positions $\vec R_i$ with the density $n^{\rm imp}$,
\begin{align}
H  =&\sum_{\bm{k}, \alpha, \beta}h_{\alpha\beta}(\bm{k})\psi_{\alpha,\bm{k}}^{\dag}\psi_{\beta,\bm{k}} +\sum_{\alpha, \beta}((\epsilon-\mu) \delta_{\alpha \beta}- \vec{B} \vec{\sigma})d_{\alpha}^{\dag}d_{\beta}\nonumber\\
&+V   \sum_{\bm{k}, \alpha,i} e^{-i \vec k \vec R_i} \psi_{\alpha,\bm{k}}^{\dag}d_{\alpha} + {\rm h.c.}
\label{ham}
\end{align}
Here the first term accounts for the motion of the surface Dirac fermions: $h_{\alpha \beta}(\bm{k})$ = $\hbar v_{F}(k_{x} \sigma_{y} - k_{y}\sigma_{x})_{\alpha \beta}-\mu \delta_{\alpha \beta}$, where $\mu$ is the electrochemical potential controlled by the gate voltage. The second term describes a localized impurity state with resonance energy $\epsilon$ which, in the third term, hybridizes with the continuum states with hybridization strength $V$. The magnetic field has already been gauged away leaving the only remaining effect as the Zeeman coupling to the impurity states. Here we use units where $g^{\rm imp} \mu_B/2=1$ where $g^{\rm imp}$ is the $g$ factor of the impurity. While our model is not expected to give a microscopic description of the actual disorder in our experiment, it is a minimal model capturing the field-induced anisotropic scattering and the interplay of magnetic (spin-flip) and non-magnetic (non-spin-flip) scattering essential to explain the experiment.

\begin{figure}
\begin{center}
\includegraphics[width=8.7cm]{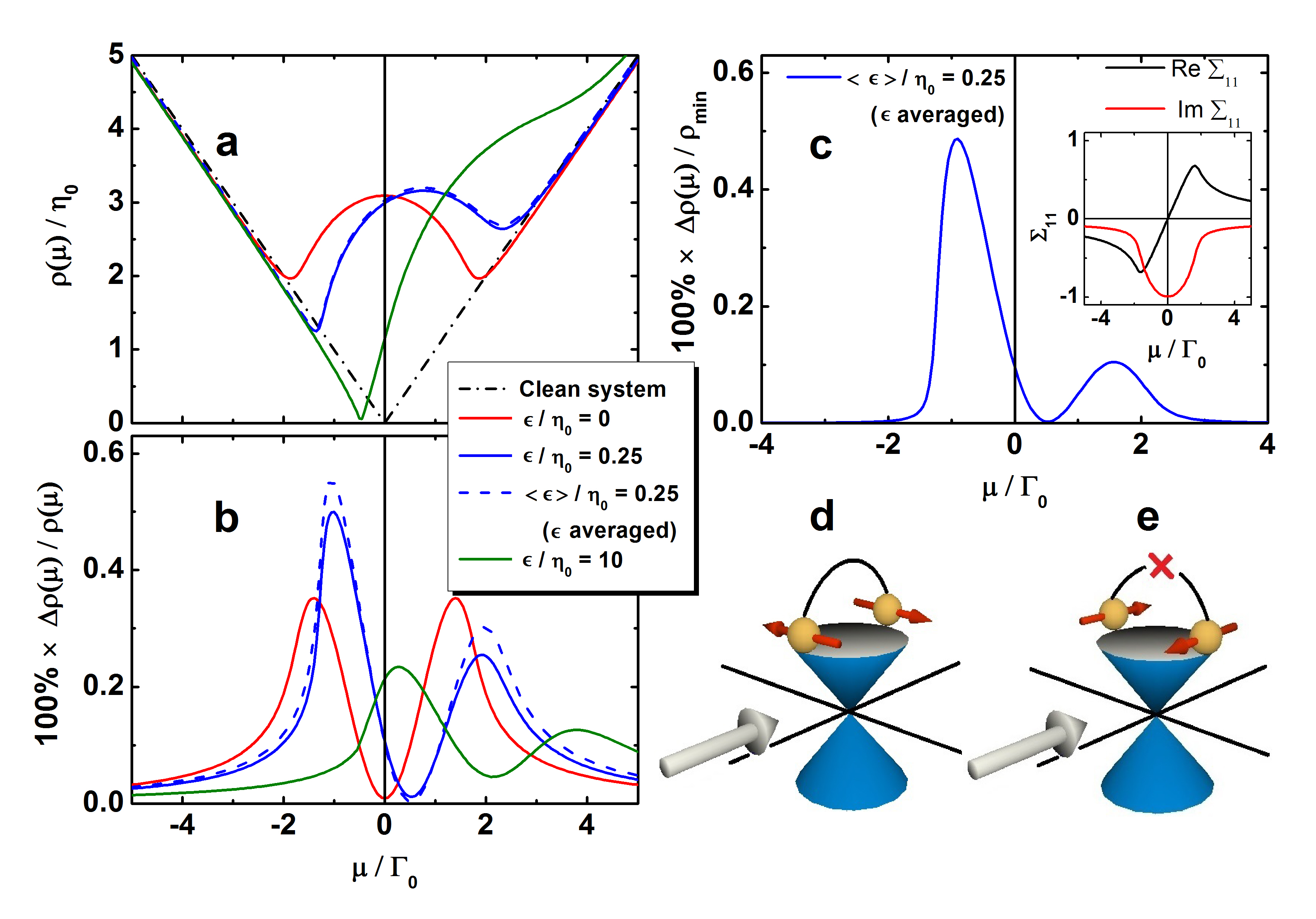}
\caption{
{\bf Origin of MR anisotropy.} (a) Local density of states for randomly distributed impurities of concentration $n^{\rm imp}$ = 0.005 as a function of normalized chemical potential $\mu/\Gamma_0$ for three different impurity-resonance energies $\epsilon$ (solid lines) and for a Gaussian distribution of $\epsilon$ with width $\eta_0 \equiv V^2 \Gamma_0/4\pi v_F^2$ centred at $\epsilon/\eta_0=2.5$ (blue dashed line). (b) The dimensionless ratio $\delta(\mu) =\frac{\rho_{\|}(\mu)-\rho_{\perp}(\mu)}{\rho_{\|}(\mu)} $ (in percent) as a function of $\mu$ for different $\epsilon$ [same as in (a)], showing a characteristic two-peak structure near the Dirac point. (c) The normalized anisotropy magnitude $\Delta \rho(\mu)/\rho_{\rm max}$ (in percent) calculated for the Gaussian distribution of $\epsilon$; this quantity is more appropriate than $\delta(\mu)$ for comparison with experiment, where bulk contributions will be present in $\rho(\mu)$. Inset shows the real and imaginary parts of the diagonal self-energy matrix element $\Sigma_{11}$ for unbroken particle-hole symmetry ($\epsilon=0$). (d), (e) Schematic picture of the scattering on {\it magnetized} impurities. (d) For Dirac fermions with spins  perpendicular to the magnetic field (gray arrow), spin-flip scattering is allowed due to  broken TRS. (e) For Dirac fermions with spins parallel/anti-parallel to the field, spin-flip scattering remains prohibited. 
} 
\label{figth}
\end{center}
\end{figure}

To calculate the conductivity $\sigma$, we employ a self-consistent T-matrix calculation \cite{Mahan, Mirlin}, valid in the limit of small $n^{\rm imp}$ for arbitrary values of $V$ and $\epsilon$.  While this approximation does not treat correctly all logarithmic corrections, it is known \cite{Mirlin} to give accurate results in situations like our experiment where weak localization or antilocalization effects are not visible. To obtain conductivities within the T-matrix approximation quantitatively, we have to include  the corresponding vertex corrections \cite{SI}, which enhance the conductivity away from the Dirac node by approximately a factor 2 (similar vertex corrections vanish for local scattering in graphene \cite{Mirlin}). For the dimensionless ratio, $\delta(\mu)=\frac{\sigma_{\perp}(\mu)-\sigma_{\|}(\mu)}{\sigma_{\|}(\mu)}= \frac{\rho_{\|}(\mu)-\rho_{\perp}(\mu)}{\rho_{\|}(\mu)}$, we find, however, that vertex corrections have only a minor effect.
 
The main result of our calculation is shown in Fig.~\ref{figth}b, where $\delta(\mu)$ is shown as a function of the chemical potential for various values of $\epsilon$. As in the experiment a clear two-peak structure emerges. The asymmetry of the two peaks is controlled by $\epsilon$ parametrizing the breaking of particle hole-symmetry by our scatterers \cite{ib_1, ib_2, CastroNeto}. The peaks track precisely the minima in the local density of states shown in Fig.~\ref{figth}a. 

Whilst we include vertex corrections in our calculation, it is instructive to consider a simplified version ignoring vertex corrections, considering only the contribution from the product of retarded and advanced Green's function \cite{Mirlin}. For $T=0$ and field $\vec B$ we obtain
\begin{widetext}
\begin{equation}
\begin{split}
\sigma^{\|/\perp}&=\frac{2 e^2 v_F^2  }{\pi} \int \frac{d^2 {\bf k}}{(2\pi)^2} \frac{|\mu-\Sigma_{11}|^2 \pm (v_F k_{\|}^2- |v_F k_{\perp}-\Sigma_{12}|^2) }{|(\mu-\Sigma_{11})^2-(v_F^2 k_{\|}^2+(v_F k_{\perp}-\Sigma_{12})^2)|^2}\approx \sigma_0 \left(1\mp \frac{c}{2} \left(\frac{{\rm Im} \Sigma_{12}}{{\rm Im} \Sigma_{11}}\right)^2\right),
\label{del}
\end{split}
\end{equation} 
\end{widetext}
where $\Sigma_{11}$ and $\Sigma_{12}$ are the diagonal (non-spin-flip) and off-diagonal (spin-flip) contributions to the self energy at $\mu$, respectively; the prefactor $c$ is numerically found to be 0.8 -- 1.2, depending on parameters. The second equality was derived by using $\Sigma_{12}\ll \Sigma_{11}$ and adsorbing the real part of $\Sigma_{12}$ by a shift of $k_y$. 
The expression of $\sigma$ in Eq. (\ref{del}) reproduces the expected behavior: 
The conductivity parallel to the magnetic field is reduced, as backscattering in parallel direction is activated by the field, as show schematically in Figs.~\ref{figth}d--e. The effect is quadratic in $B$ for small $B$ as ${\rm Im} \Sigma_{12}$, the spin-flip scattering rate, is linear in $B$ for small $B$.

According to Eq. (\ref{del}), the peaks in the anisotropic resistivity arise from peaks in ${\rm Im} \Sigma_{12}$. As shown in the Supplementary Information \cite{SI},
 ${\rm Im}\, \Sigma_{12} \propto  {\rm Im}[ (\Sigma_{11})^2] =2\, {\rm Im}\, \Sigma_{11}\,{\rm Re} \,\Sigma_{11} $. Using this result, we find that the origin of the peak can be traced back to peaks in ${\rm Re} \,\Sigma_{11}$. Due to the Kramers-Kr\"onig relation, these peaks occur when $|{\rm Im}\, \Sigma_{11}|$ quickly diminishes as $\mu$ moves away from the Dirac point (see Fig. 3c inset); this diminishment occurs roughly at $\mu \sim \pm \Gamma_0$, where $\Gamma_0=- {\rm Im}\, \Sigma_{11}(\mu=0)$ is the non-spin-flip scattering rate at the Dirac point. Hence, the location of $\mu$ associated with a peak gives a measure of $\Gamma_0$. Estimating $\mu$ from the Hall data assuming a Fermi velocity of $3.9\times 10^5$\,m/s \cite{Band_engin, VF_BST} (see Fig. 2d inset), we obtain for our sample $\Gamma_0 \sim 50\,$meV, corresponding to a mean free path $v_F/\Gamma_0$ of $\sim 70$\,\AA. 
 
 
Furthermore, according to Eq. (\ref{del}), the amplitude of the anisotropy is set by the square of the ratio of spin-flip and non-spin-flip scattering rates. Since our data show $\delta \sim 1\% $ at 9 T, one may infer that about 10\% of the scattering processes are spin-flipping at 9 T and the corresponding spin-flip scattering rate and mean free path are $\sim$5 meV and $\sim$700\,\AA, respectively. 

It is important to note that the two-peak structure in the anisotropy and its relation to the microscopic parameters are found to be robust even when we consider distributions of $V$ and $\epsilon$. As an example, Fig.~\ref{figth}b compares the results for the cases when $\epsilon$ is fixed or has a Gaussian distribution (the result is similar for a Gaussian distribution of $V$ \cite{SI}). Note, however, that microscopic details will strongly affect the asymmetry of the peaks. The experimental data for the PHE amplitude (Fig. 2d) is best compared to Fig. \ref{figth}c.

The proposed theoretical model gives a clear physical picture for the origin of the MR anisotropy in the TI surface; namely, the spin-momentum locking protects the surface Dirac fermions from backscattering in zero field, but the in-plane magnetic field breaks TRS and magnetizes randomly distributed impurities, drastically changing this topological protection. For spins perpendicular to the field, the protection is lifted and backscattering is allowed due to TRS breaking as schematically shown in Fig.~\ref{figth}d. In contrast, for spins parallel or anti-parallel to the field [shown in Fig.~\ref{figth}e], backscattering is still forbidden. This anisotropy in the scattering rate directly results in the AMR and PHE: For the field direction parallel to the current (and hence perpendicular the spin orientation), the allowed backscattering results in the increased resistance $R_{||}$, whilst for the field direction perpendicular to the current, the resistance $R_{\perp}$ will be much less affected. This is the reason for a positive anisotropy amplitude (i.e. $R_{||}-R_{\perp} \geq$ 0). We further show that, from the two-peak structure of the anisotropy, one can infer the effective scattering rates for both spin-flip and non-spin-flip scatterings. Therefore, the PHE discovered here represents a novel signature of TRS breaking in TIs and provides microscopic information on the topological surface transport.

\begin{flushleft} 
{\bf Methods}

{\bf MBE Growth of high-quality BST films.} Bi$_{2−x}$Sb$_{x}$Te$_{3}$ films with the thickness in the range of 11--17 nm were grown on sapphire (0001) substrates by co-evaporation of high-purity Bi, Sb, and Te from Knudsen cells in the ultra-high vacuum MBE chamber. The flux ratio of Bi and Sb was optimized for obtaining most bulk-insulating films and was kept at $1:5.5$. The Te flux exceeded the aggregated flux of Bi and Sb by at least 10 times. The deposition was done in three temperature steps: at 230$^{\circ}$C for 5 min, at 280$^{\circ}$C for 5 min, and finally at 325$^{\circ}$C for a time period which is sufficient to grow a film with a desirable thickness. The thickness and morphology of the grown films were measured {\it ex situ} by AFM.

{\bf Device microfabrication.} To make a dual-gate device, the grown films need to be transferred from the sapphire substrate to a Si/SiO$_{2}$ wafer, which serves as a back-gate electrode and dielectric. The separation of a BST film from the substrate was done by first spin-coating the film with PPMA and then dipping into 5\% KOH aqueous solution to initiate the detachment. The full detachment was done by slowly dipping the film into distilled water. The detached BST/PMMA bilayer was fished out on the Si/SiO$_{2}$ wafer, dried at room temperature, treated with acetone to remove PMMA, and annealed at 120$^{\circ}$C for several hours under vacuum conditions to remove residual water. To pattern the BST film into a Hall bar, we employed photolithography. Exposed parts of the film were etched out in HCl/H$_{2}$O$_{2}$/CH$_{3}$COOH aqueous solution. As the top-gate dielectric, 200-nm-thick SiN$_{x}$ layer was deposited by using hot-wire CVD at temperatures below 80$^{\circ}$C. The top-gate electrode and metal contact pads were made by Ti/Au deposition.

{\bf Magneto-resistivity measurements.} Both {\it ac} and {\it dc} techniques were employed for resistivity and Hall-effect measurements. The top- and bottom-gate voltages were controlled by two independent Keithley 2450 source meters. A single-axis rotation probe with a capability of mounting the sample horizontally or vertically was used for both out-of-plane and in-plane rotations in magnetic fields. Special care has been taken to isolate the genuine in-plane magnetic-field effects from spurious contributions due to a possible misalignment of the sample \cite{SI}.

{\bf Theoretical calculations.} A self-consistent T-matrix approach was employed to calculate the self-energy of the Dirac electrons. The Kubo formula  was used to calculate the conductivity within this approximation, including the appropriate vertex corrections, see Supplementary Material for details \cite{SI}.

\end{flushleft}

\begin{flushleft} 
{\bf Acknowledgements: }
This work was supported by DFG (CRC1238 ``Control and Dynamics of Quantum Materials", Projects A04, B01, and C02).
\end{flushleft}

\begin{flushleft} 
{\bf Author contributions:}
A.A.T. and Y.A. conceived the project. A.A.T. and S.S. grew the topological insulator thin films. F.Y., Y.K., and K.M. made the devices. A.A.T. performed measurements and analyzed the data. H.L performed the theoretical calculations, H.L. and A.R.  developed the interpretation. A.A.T., H.L., A.R., and Y.A. wrote the manuscript with inputs from all authors.
\end{flushleft} 

\begin{flushleft} 
{\bf Additional information}
Correspondence and requests for materials should be addressed to Y.A. (ando@ph2.uni-koeln.de).
\end{flushleft} 

\clearpage
\onecolumngrid

\renewcommand{\thefigure}{S\arabic{figure}} 

\setcounter{figure}{0}

\begin{flushleft} 
{\Large {\bf Supplemental Material}}
\end{flushleft} 
\vspace{2mm}
\begin{flushleft} 
{\bf S1. Angular dependences of AMR and PHE}
\end{flushleft} 

When a resistivity anisotropy is induced by an in-plane magnetic field, the resistivity tensor may be written in a diagonalized form by taking the magnetic-field direction as the $x'$ axis of the principal coordinates:
\begin{equation}
\left( {\begin{array}{c}
E_{x'}\\
E_{y'}\\ 
\end{array} } \right) = 
\left( {\begin{array}{cc}
R_{\parallel} & 0 \\
0 & R_{\perp} \\
\end{array} } \right) 
\left( {\begin{array}{c}
j_{x'}\\
j_{y'}\\ 
\end{array} } \right).
\end{equation}
Here, $E_{x'}$ and $j_{x'}$ are along the magnetic field, and $E_{y'}$ and $j_{y'}$ are perpendicular to the magnetic field.
When one transforms this into the coordinate system fixed on the sample, in which $x$ is the current direction and $y$ is the transverse direction on the film plane, the resistivity tensor becomes
\begin{eqnarray}
\left( {\begin{array}{c}
E_{x}\\
E_{y}\\ 
\end{array} } \right) & = &
\left( {\begin{array}{cc}
\rm{cos} \varphi & - \rm{sin} \varphi\\
\rm{sin} \varphi & \rm{cos} \varphi \\
\end{array} } \right)
\left( {\begin{array}{cc}
R_{\parallel} & 0 \\
0 & R_{\perp} \\
\end{array} } \right) 
\left( {\begin{array}{cc}
\rm{cos} \varphi & \rm{sin} \varphi\\
- \rm{sin} \varphi & \rm{cos} \varphi \\
\end{array} } \right)
\left( {\begin{array}{c}
j_{x}\\
j_{y}\\ 
\end{array} } \right),
\nonumber \\
&=& \left( {\begin{array}{cc}
R_{\parallel} \cos^2 \varphi + R_{\perp} \sin^2 \varphi &\,\,\, ( R_{\parallel} - R_{\perp}) \rm{cos} \varphi \, \rm{sin} \varphi \\
( R_{\parallel} - R_{\perp}) \rm{cos} \varphi \, \rm{sin} \varphi &\,\,\, R_{\parallel} \sin^2 \varphi + R_{\perp} \cos^2 \varphi \\
\end{array} } \right) 
\left( {\begin{array}{c}
j_{x}\\
j_{y}\\ 
\end{array} } \right).
\end{eqnarray}

By setting $j_{y}$ = 0 as the boundary condition to represent our measurement configuration, one obtains
\begin{equation}
R_{xx} = E_{x}/j_{x} = R_{\perp} + ( R_{\parallel} - R_{\perp}) \rm{cos} ^2 \varphi,
\end{equation}
and
\begin{equation}
R_{yx} = E_{y}/j_{x} = ( R_{\parallel} - R_{\perp}) \rm{cos} \varphi \, \rm{sin} \varphi .
\end{equation}
Here $R_{yx}$ represents the planar Hall effect (PHE), which is essentially an off-diagonal component of the in-plane magnetoresistance. An important difference from the ordinary Hall effect is that this component is symmetric with respect to the magnetic field, as is actually observed in our TI devices (see Fig. S1 inset). The anisotropic magnetoresistance (AMR) manifests itself in $R_{xx}$.

\begin{figure}
\begin{center}
\includegraphics[height=9cm]{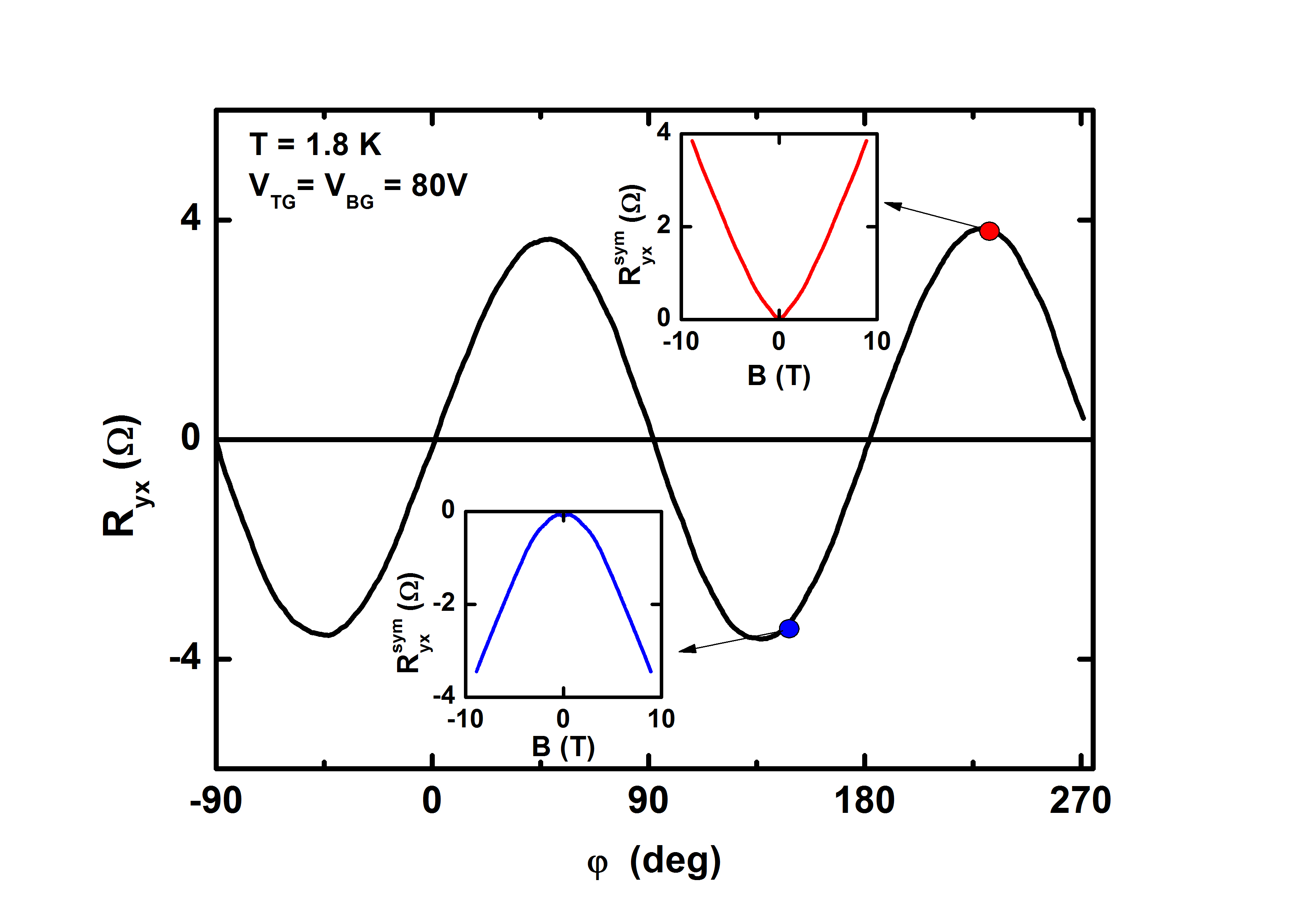}
\caption{
Angular dependence of the planar Hall effect (PHE) measured in one of the 17-nm-thick device at 1.8 K, $B$ = 9 T, and $V_{TG} = V_{BG}$ = 80 V. The angle $\varphi$ is defined in Fig. 2 of the main text. Insets show the magnetic-filed dependence of the symmetric part of $R_{yx}$ measured at two different angles $\varphi$.
} 
\end{center}
\end{figure}

\begin{flushleft} 
{\bf S2. Effect of sample misalignment \\
(Spurious contribution to the angular dependence)}
\end{flushleft} 

{\bf Arbitrary rotation:} 
An arbitrary 3D rotation is specified by an axis of rotation together with an angle of rotation about this axis (one also needs to specify the orientation of the axis and whether the rotation is taken to be clockwise or counterclockwise with respect to this orientation).
A counterclockwise rotation about an arbitrary unit vector $\vec u = (u_{x}, u_{y}, u_{z})$ by angle $\psi$ is given by the transformation matrix
\begin{equation}
R_{\vec u}(\psi)=
\left( {\begin{array}{ccc}
\cos \psi + u_{x}^{2}(1-\rm{cos}\psi) &u_{x}u_{y}(1-\cos\psi) - u_{z}\sin\psi\,\,\,\,& u_{x}u_{z}(1-\cos\psi) + u_{y}\sin\psi\\
u_{y}u_{x}(1-\cos \psi) + u_{z}\sin\psi & \cos\psi + u_{y}^{2}(1-\cos\psi)& u_{y}u_{z}(1-\cos\psi) - u_{x}\sin\psi\\
u_{z}u_{x}(1-\cos \psi) - u_{y}\sin\psi \,\,\,\,&u_{z}u_{y}(1-\cos\psi) + u_{x}\sin\psi& \cos\psi + u_{z}^{2}(1-\cos\psi)\\
\end{array} } \right).
\end{equation}
A change of the reference frame can be quantified by a rotation about a suitable axis. In this respect, if two frames are related by a rotation about the unit vector $\vec u$ by angle $\psi$, a vector $\vec a = (x, y, z)$ in the original frame is expressed in the new frame as
\begin{equation}
\left( {\begin{array}{c}
x'\\
y' \\
z' \\
\end{array} } \right) 
= R_{\vec u}(\psi)
\left( {\begin{array}{c}
x\\
y \\
z \\
\end{array} } \right).
\label{eq:rot}
\end{equation}

{\bf Out-of-plane magnetic field due to the misalignment:}
Now we consider the situation of our experiment to rotate the magnetic field in the film plane. In the actual experiment, the magnetic field is fixed in the $z$ axis of the laboratory frame, i.e. $\vec B$ = $(0, 0, B)$, and the rotation is performed with a mechanical rotator, which rotates the sample around the $y$ axis of the laboratory frame. We assume that, before any rotation (i.e. rotation angle $\psi$ = 0), the current is along the $x$ axis of the laboratory frame. 

Ideally, for the in-plane rotation, the $x'y'$ plane of the sample frame should be identical to the $zx$ plane of the laboratory frame. In reality, however, there is some misalignment of the sample on the rotator, which results in the deviation of the $z'$ axis of the sample frame from the actual rotation axis ($y$ axis of the laboratory frame). This deviation can be parametrized by using two misalignment angles; namely, rotations of the $y$ axis of the laboratory frame by angles $\delta$ and $\alpha$ around the $x'$ and $y'$ axes of the sample frame, respectively. After these two rotations, the $y$ axis of the laboratory frame is brought to the $z'$ axis of the misaligned sample frame.

By using Eq. (\ref{eq:rot}), the unit vector of the rotation axis [which is $\vec e_{y} = (0, 1, 0)$ in the laboratory frame] is expressed in the sample frame as
\begin{equation}
\vec {u'} = \left( {\begin{array}{c}
u_{ x'}\\
u_{y'} \\
u_{z'} \\
\end{array} } \right) = 
\left( {\begin{array}{ccc}
\rm{cos} \alpha & 0 & -\rm{sin} \alpha \\
0 & 1 & 0 \\
\rm{sin} \alpha & 0 & \rm{cos} \alpha \\
\end{array} } \right) 
\left( {\begin{array}{ccc}
1 & 0 & 0 \\
0 & \rm{sin}\delta & \rm{cos}\delta \\
0 & -\rm{cos}\delta & \rm{sin}\delta \\
\end{array} } \right)
\left( {\begin{array}{c}
0\\
1 \\
0 \\
\end{array} } \right) =
\left( {\begin{array}{c}
\rm{cos}\delta \, \rm{sin}\alpha \\
\rm{sin}\delta \\
- \rm{cos}\delta \, \rm{cos}\alpha \\
\end{array} } \right) .
\label{eq:u'}
\end{equation}
Also, the magnetic-field vector in the sample frame before the sample rotation ($\psi$ = 0) is
\begin{equation}
\vec {B'}(0) = \left( {\begin{array}{c}
B_{x'}(0)\\
B_{y'}(0) \\
B_{z'}(0) \\
\end{array} } \right) = B 
\left( {\begin{array}{ccc}
\rm{cos} \alpha & 0 & -\rm{sin} \alpha \\
0 & 1 & 0 \\
\rm{sin} \alpha & 0 & \rm{cos} \alpha \\
\end{array} } \right) 
\left( {\begin{array}{ccc}
1 & 0 & 0 \\
0 & \rm{sin}\delta & \rm{cos}\delta \\
0 & -\rm{cos}\delta & \rm{sin}\delta \\
\end{array} } \right)
\left( {\begin{array}{c}
0\\
0 \\
1 \\
\end{array} } \right) = B
\left( {\begin{array}{c}
- \rm{sin}\delta \, \rm{sin}\alpha \\
\rm{cos}\delta \\
\rm{sin}\delta \, \rm{cos}\alpha \\
\end{array} } \right).
\end{equation}
Now, when the sample is rotated clockwise by angle $\psi$ around the axis $\vec {u'}$ obtained in Eq. (\ref{eq:u'}), the rotation matrix $ R_{\vec {u'}}(\psi)$ is
\begin{equation}
\scriptsize{ \left( {\begin{array}{ccc}
\rm{cos} \psi +\rm{cos}^{2}\delta \, \rm{sin}^{2} \alpha (1-\rm{cos}\psi) & \rm{cos}\delta \, \rm{sin}\delta \, \rm{sin}\alpha \,(1-\rm{cos}\psi) - \rm{cos}\delta \, \rm{cos}\alpha \,\rm{sin}\psi & - \rm{cos}^{2}\delta \, \rm{cos}\alpha \, \rm{sin}\alpha \,(1-\rm{cos}\psi) - \rm{sin}\delta \, \rm{sin}\psi\\
\\
\rm{cos}\delta \, \rm{sin}\delta \, \rm{sin}\alpha \,(1-\rm{cos}\psi) + \rm{cos}\delta \, \rm{cos}\alpha \,\rm{sin}\psi & \rm{cos} \psi +\rm{sin}^{2}\delta \, (1-\rm{cos}\psi)& - \rm{cos}\delta \, \rm{sin}\delta \, \rm{cos}\alpha \,(1-\rm{cos}\psi) + \rm{cos}\delta \, \rm{sin}\alpha \,\rm{sin}\psi\\
\\
- \rm{cos}^{2}\delta \, \rm{cos}\alpha \, \rm{sin}\alpha \,(1-\rm{cos}\psi) + \rm{sin}\delta \, \rm{sin}\psi & - \rm{cos}\delta \, \rm{sin}\delta \, \rm{cos}\alpha \,(1-\rm{cos}\psi) - \rm{cos}\delta \, \rm{sin}\alpha \,\rm{sin}\psi & \rm{cos} \psi +\rm{cos}^{2}\delta \, \rm{cos}^{2} \alpha (1-\rm{cos}\psi)\\
\end{array} } \right)}.
\end{equation}
Finally, the magnetic-field vector in the sample frame after the sample is rotated clockwise about the $\vec {u'}$ axis by angle $\psi$ is written as
\begin{equation}
\vec {B'}(\psi) = \left( {\begin{array}{c}
B_{x'}(\psi)\\
B_{y'}(\psi) \\
B_{z'}(\psi) \\
\end{array} } \right) = B 
\left( {\begin{array}{c}
- \rm{cos} \alpha \, \rm{sin} \psi \, - \, \rm{sin} \delta \, \rm{sin} \alpha \, \rm{cos} \psi\\
\rm{cos} \delta \, \rm{cos} \psi \\
\rm{sin} \delta \, \rm{cos} \alpha \, \rm{cos} \psi \, - \, \rm{sin} \alpha \, \rm{sin} \psi \\
\end{array} } \right) .
\end{equation}
The magnetic-field component perpendicular to film plane, $B_{z'}(\psi)$, can be written as
\begin{equation}
B_{z'}(\psi) = B\, \sqrt{ \rm{sin}^{2} \delta \, \rm{cos}^{2} \alpha \, + \rm{sin}^2 \alpha \, }\,\, \rm{cos} (\psi \,+\, \phi), \,\,\,\,\,
\phi = \rm{arctan}\left(\frac{ \rm{sin}\, \alpha}{ \rm{sin} \delta \, \rm{cos} \alpha }\right).
\end{equation}
In our experiment, due to the design of the rotating sample stage, the condition $\alpha$ $\ll$ $\delta$ $\ll$1 holds and one may obtain
\begin{equation}
B_{z'}(\psi) \simeq B\, \rm{sin} \delta \, \rm{cos} \psi.
\end{equation}

For $B$ = 9 T and $\delta$ as small as 0.5$^{\circ}$, the magnetic-field component perpendicular to the sample surface due to the misalignment 
would be maximally $B\, \sin \delta \approx$ 78.5 mT (785 G). 
The ordinary orbital MR (expected for magnetic fields perpendicular to the surface) has a quadratic field dependence in this magnetic-field range, and hence its $\psi$ dependence would be $\sim \rm{cos} ^2 \psi$, which is indistinguishable from the AMR behaviour. 
On the other hand, the ordinary Hall resistivity, caused by a misalignment, has a linear magnetic-filed dependence and has a completely different $\psi$ dependence ($\sim \rm{cos} \psi$) than PHE.

\begin{figure}[b]
\begin{center}
\includegraphics[height=8cm]{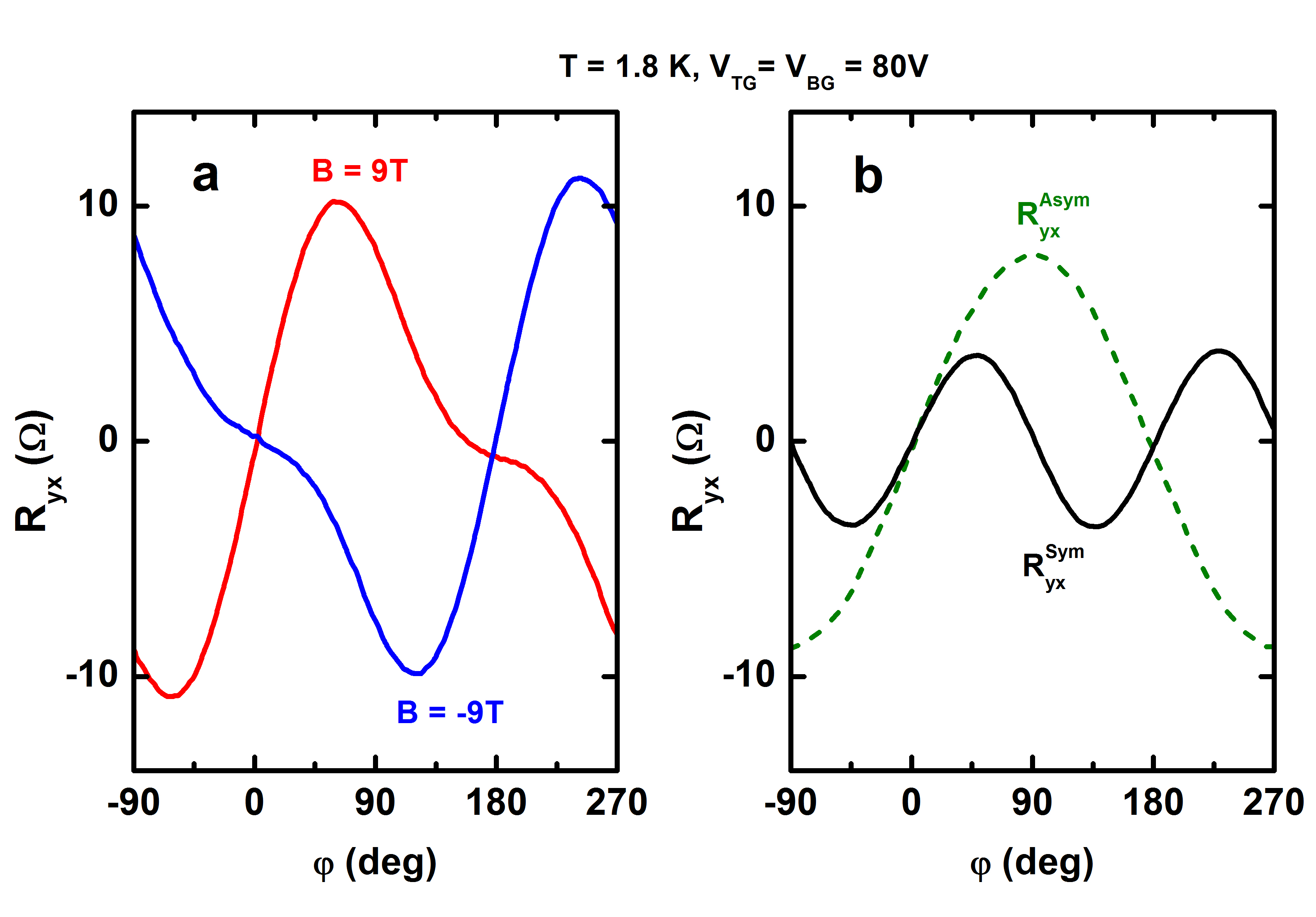}
\caption{
(a) Raw $R_{yx}$ data from a 17-nm device at $\mid B \mid$ = 9 T and $V_{TG} = V_{BG}$ = 80 V, and (b) the decomposition into symmetric and antisymmetric parts. The misalignment is about 1.5$^{\circ}$.
} 
\end{center}
\end{figure}

\begin{figure}
\begin{center}
\includegraphics[height=11.5cm]{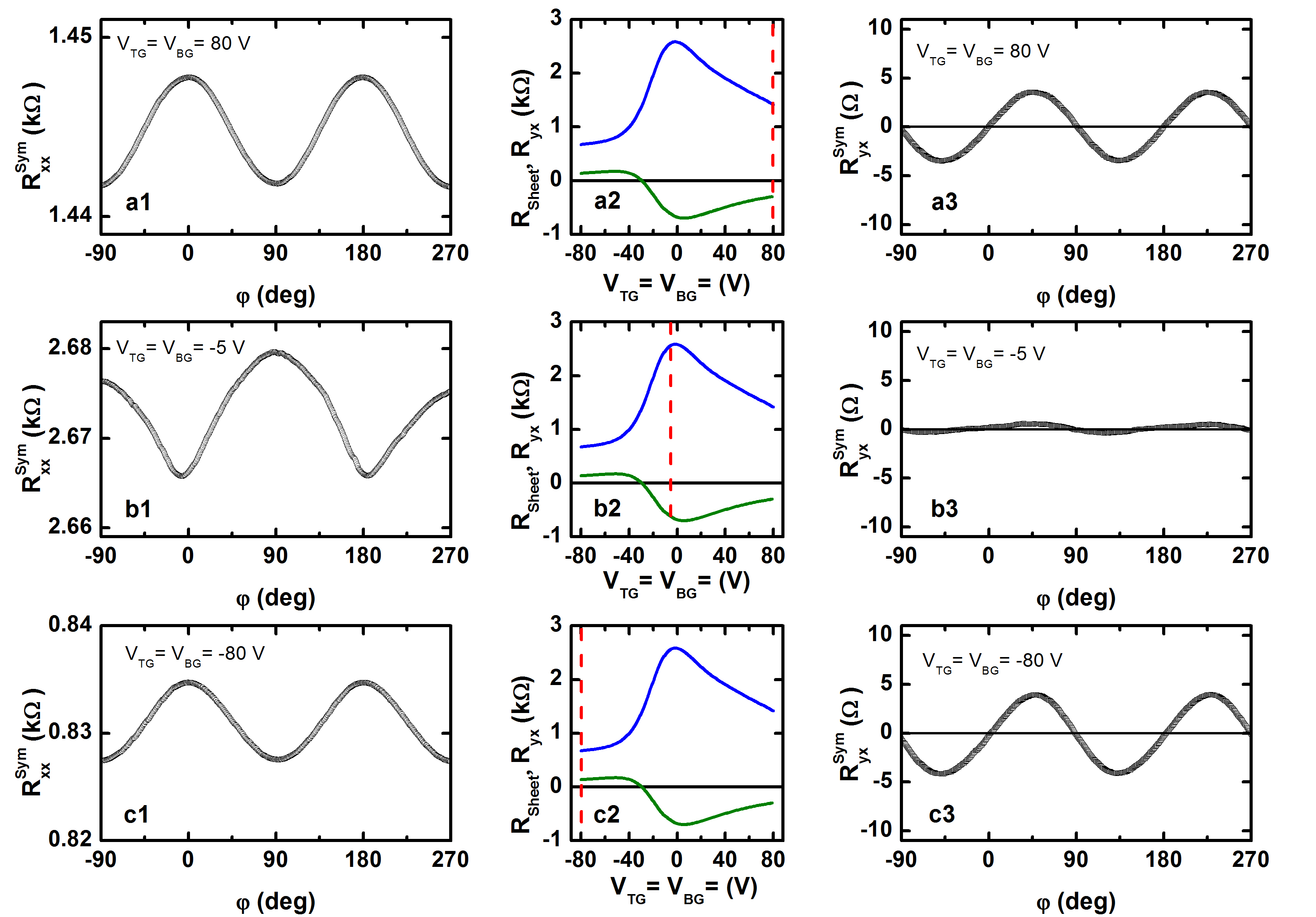}
\caption{
Comparison of the AMR and PHE behaviors at different $V_{TG}=V_{BG}$ in $B$ = 9 T. The middle panels show the corresponding $R_{xx}$(0T) and $R_{yx}$(9T) data measured for the out-of-plane configuration of the magnetic field.
} 
\end{center}
\end{figure}

\begin{figure}[htpb]
\begin{center}
\includegraphics[height=7.5cm]{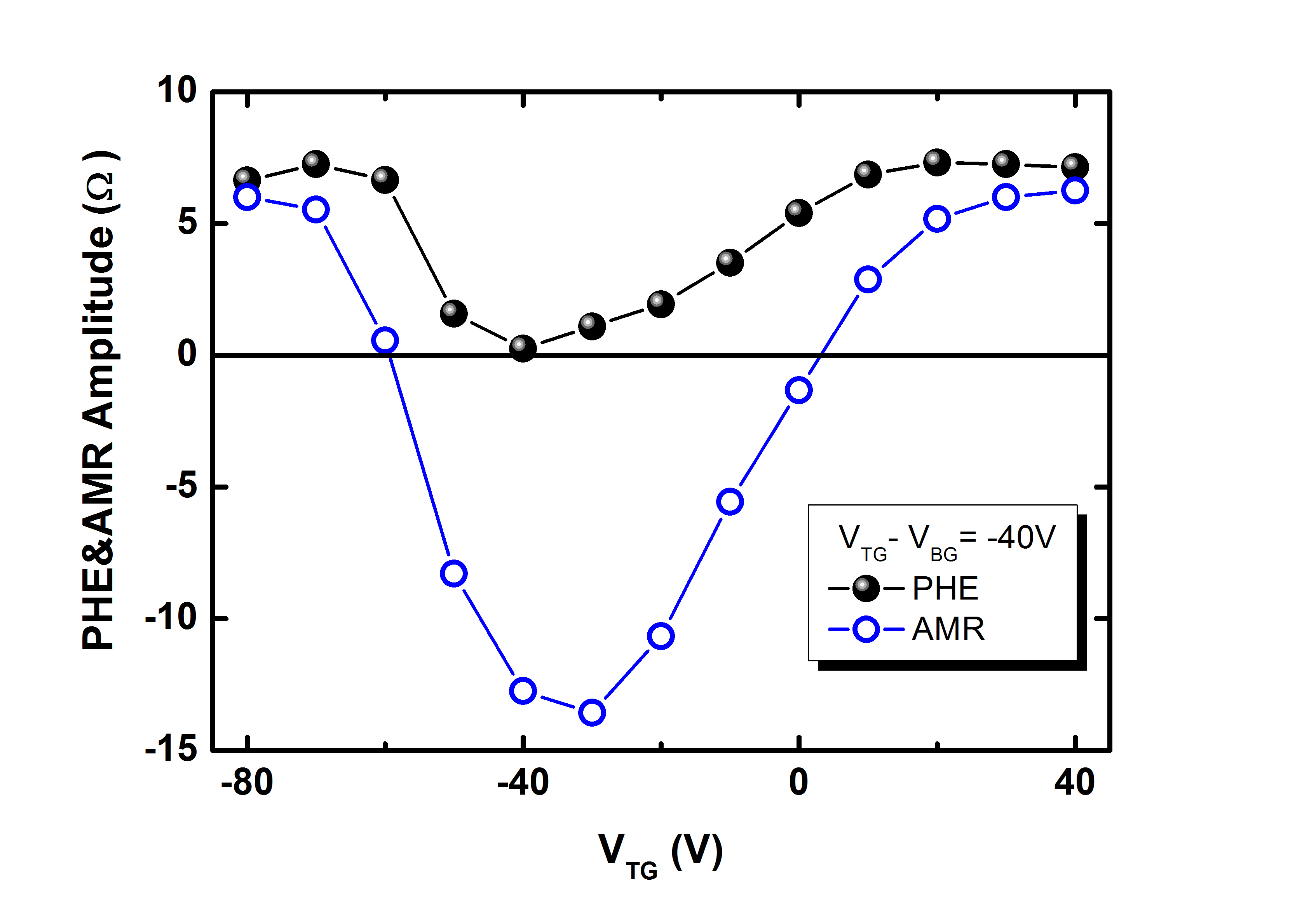}
\caption{
Comparison of the gate-voltage dependences of the amplitudes of AMR and PHE, measured along the gating path with $V_{TG}-V_{BG}$ = $-40$ V in $B$ = 9 T.
} 
\end{center}
\end{figure}

\begin{figure}
\begin{center}
\includegraphics[height=7.5cm]{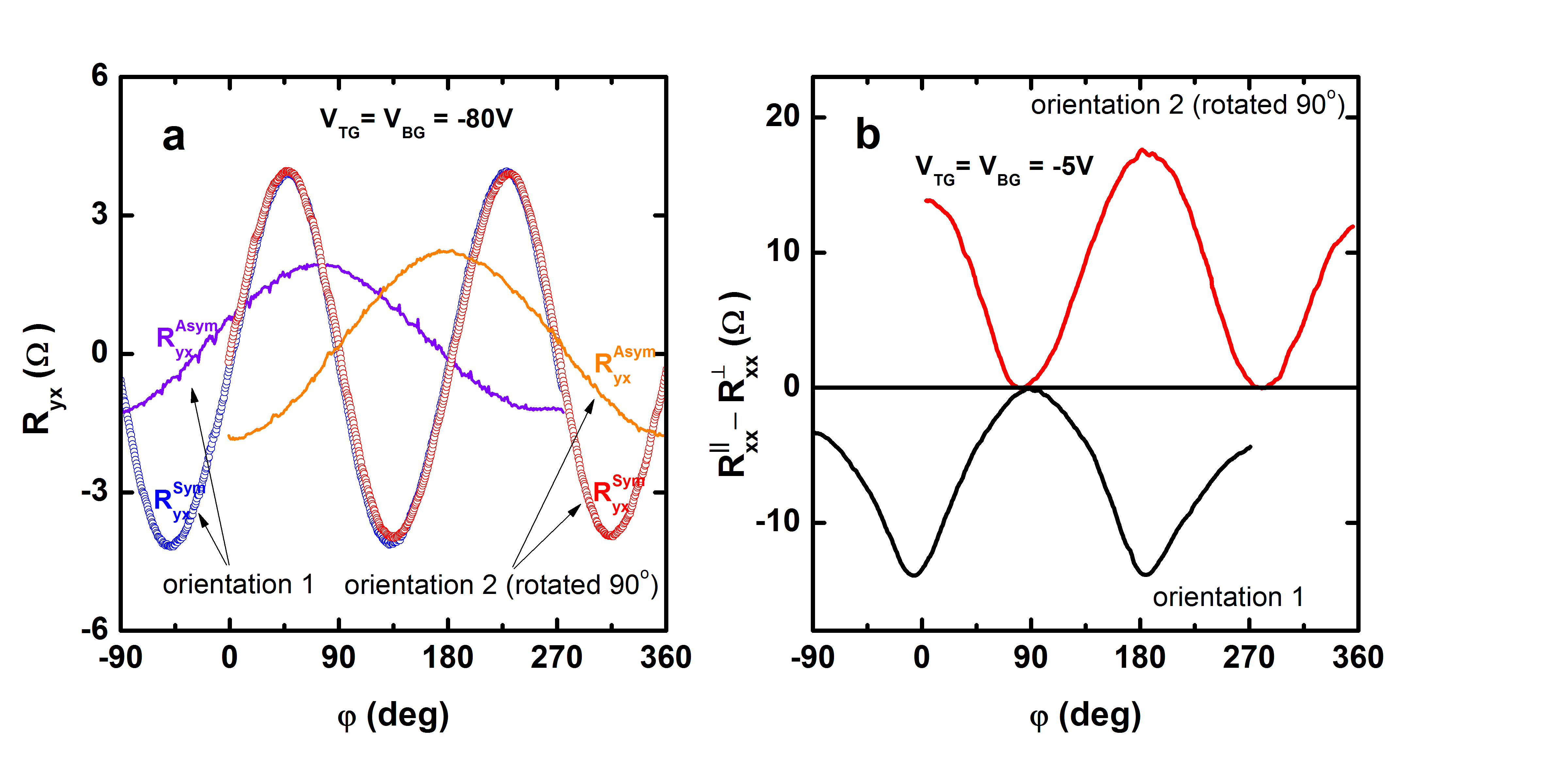}
\caption{
(a) PHE and (b) AMR in $B$ = 9 T for two different mounting orientations of the sample. 
} 
\end{center}
\end{figure}

\begin{figure}[!b]
\begin{center}
\includegraphics[height=12cm]{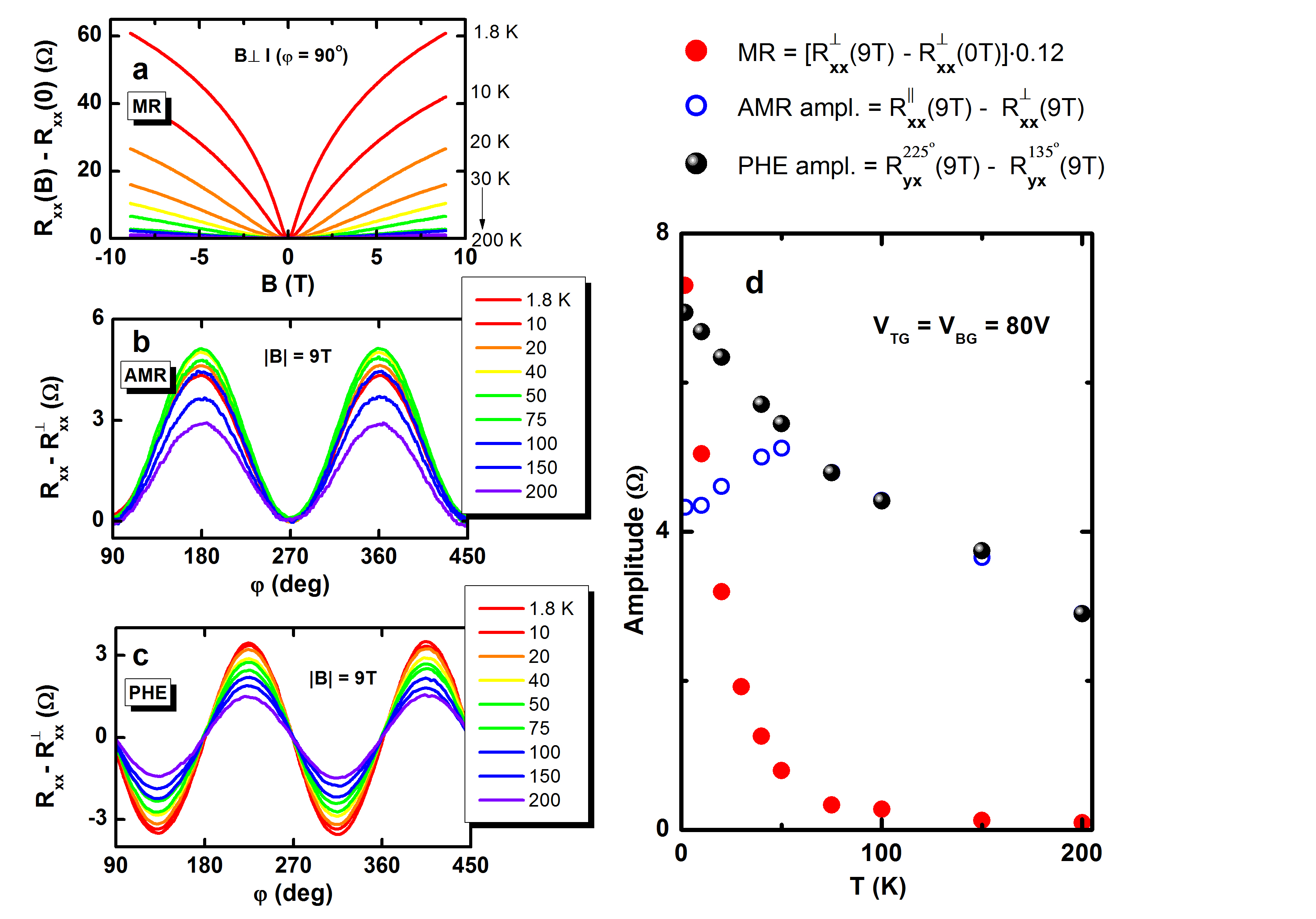}
\caption{
(a) MR, (b) AMR, and (c) PHE measured at different temperatures on a 17-nm-thick device at $V_{TG} = V_{BG}$ = 80 V. (d) Temperature dependences of their amplitudes at $B$ = 9 T.
} 
\end{center}
\end{figure}

\begin{flushleft} 
{\bf S3. Evidence in the experimental data for sample misalignment}
\end{flushleft} 

{\bf Planar Hall Effect:}
Figure S2(a) shows the raw $R_{yx}$ data measured in 9 T and $-9$ T in one of 17-nm-thick devices with an estimated misalignment angel $\delta$ of $-1.5^{\circ}$. The decomposition of $R_{yx}$ into symmetric and antisymmetric parts is shown in Fig. S2(b). The symmetric part [black solid line in Fig. S2(b)] is the PHE signal, which follows $\sim$cos$\varphi \,$sin$\varphi$ dependence. Its amplitude does not depend on a misalignment angle or a mounting configuration. The antisymmetric part (green dashed line) is an ordinary Hall contribution, which here follows $\sim$sin$\varphi$ dependence. Its amplitude depends on a misalignment angle $\delta$ as $\sim$sin$\delta$. The phase is not universal and depends on a mounting configuration as will be shown below.

\vskip 0.3cm
{\bf Spurious ``Negative" AMR upon gating:}
Figure S3 shows the AMR behavior (left column), the gate-voltage dependences of $R_{xx}$(0T) and $R_{yx}$(9T) (central column), and the PHE behavior (right column), measured in one of the 17-nm-thick devices with a dc current of 30 $\mu$A. The AMR and PHE data were taken at three different gate voltages as indicated in the central column by vertical dashed lines. Let us first consider the development of the AMR; it is positive (i.e. $R_{\parallel} > R_{\perp}$) at $V_{TG}$ = $V_{BG}$ = 80 V, when the charge carriers on both surfaces are electrons. For $V_{TG}$ = $V_{BG}$ = $-5$ V, when the Fermi level is close to the Dirac point and $R_{xx}$ is maximal, the observed AMR is negative, similar to what was observed in Ref. \cite{Sulaev_Nano}. Since we have top and bottom gates (and thinner samples), we can move the Fermi level even further into the $p$-doped side in comparison to the measurements in Ref. \cite{Sulaev_Nano}. At $V_{TG}$ = $V_{BG}$ = $-80$ V, when the charge carriers on both surfaces are holes, the sign of AMR changed again. If we take a look at the PHE, its amplitude never changes sign (although it becomes close to zero near the Dirac point). 

Figure S4 shows the gate-voltage dependences of the amplitudes of the AMR and PHE measured along another gating path, in which we kept $V_{TG}-V_{BG}$ = $-40$ V. Here again, the AMR shows a sign change, while the PHE amplitude remains positive. 

As has been discussed in Sec. S1, the contribution from the orbital MR due to a misalignment is difficult to distinguish from the genuine AMR signal in the $R_{xx}$ data. Therefore, it is reasonable to assume that the difference in the amplitudes of the AMR and PHE comes from the finite contribution of the orbital MR to AMR, which can give rise to a negative total signal when the MR due to the out-of-plane field is large.

To test this assumption, we performed the following experiment: We measured the same device twice with two different orientations of the sample on the same sample holder. The orientations differ from each other by 90$^{\circ}$ rotation along the axis perpendicular to the sample surface. The genuine signal should not depend on the orientation of the sample mounting, while for the signal coming from a misalignment, its phase should shift by 90$^{\circ}$. As can be seen in Fig. S5(a), the symmetric parts of $R_{yx}$ (i.e. the genuine PHE signals) were absolutely the same in both measurements. The antisymmetric parts (i.e. ordinary Hall contributions) in the two measurements were indeed shifted by 90$^{\circ}$ relative to each other. In the results for $R_{xx}$ shown in Fig. S5(b), the phase also shifts by 90$^{\circ}$ and the amplitude changes from negative to positive, indicating that a major part of the signal is coming from the misalignment and the negative amplitude of the AMR in this case is an artifact.

\vskip 0.3cm
{\bf Temperature dependences of MR, AMR, and PHE:}
The temperature dependences of the MR, AMR, and PHE are also useful for distinguishing the genuine signal from spurious contributions. 
Figure S6 shows an example of such an examination, which was made for $V_{TG}=V_{BG}$ = 80 V, when both surfaces are filled with electrons.
It turns out that the genuine magnetic-field-induced in-plane anisotropy is not very sensitive to temperature, because the amplitude of the PHE decreases only by a factor of two upon raising the temperature from 1.8 to 200 K. On the other hand, the magnitude of the MR in the out-of-plane magnetic field of 9 T drops by a factor of 10 at 50 K [see Fig. S6(d)]. The temperature dependence of the AMR amplitude is most unusual: It initially increases with increasing temperature, reaches a maximum, and then decreases after merging with the PHE amplitude [Fig. S6(d)]. A misalignment can easily explain this behaviour: At low temperature, the contribution from the orbital MR (which appears to be negative here) is the largest. This contribution rapidly diminishes with increasing temperature, leading to an apparent increase in the AMR amplitude. At about 50 K, when the spurious contribution from the orbital MR becomes negligible, the AMR amplitude reaches its maximum and becomes identical to the PHE amplitude, as is expected from the resistivity-tensor phenomenology.


\begin{flushleft} 
{\bf S4. Symmetries and topological protection in the presence of an in-plane magnetic field}
\end{flushleft}

An external in-plane magnetic field breaks time-reversal symmetry and allows for backscattering of electrons at the surface, therefore partially lifting the topological protection of the material.
Further crystalline symmetries which exist on average even in a disordered sample can, however, guarantee the existence of gapless surface states for high-symmetry surfaces \cite{classification}.
In our experiment, we consider the (111) surface with rhombohedral R$\bar3$m symmetry. For a field in the $[1\bar{1}0]$ direction and equivalent directions obtained by $60^\circ$ rotations around the surface normal, a mirror symmetry guarantees that no average magnetization is generated perpendicular to the surface. We therefore expect that the system remains gapless for these specific field directions. More precisely, the symmetry is only present on average but also this is sufficient to stabilize a metallic surface state \cite{classification}.

For other field directions parallel to the surface, by symmetry the formation of a gapful quantum Hall state is possible and expected to happen for $T=0$ in infinitely large samples. However, for practical purposes this effect is suppressed as the magnetization perpendicular to the surface is (by symmetry) proportional to $B_\|^3$. As $B_\|$ is smaller than all relevant microscopic energy scales, only a very small effect is expected.

\begin{flushleft} 
{\bf S5. Self-consistent T-matrix approximation (SCTMA)}
\end{flushleft} 

To model scattering from a random magnetic field we consider two-dimensional Dirac electrons
coupled to impurities located at random positions $\vec R_i$ with the density $n^{\rm imp}$ (as in the main text):
\begin{equation}
H = \sum_{\bm{k}, \alpha, \beta}h_{\alpha\beta}(\bm{k})\psi_{\alpha}^{\dag}(\bm{k})\psi_{\beta}(\bm{k}) + \sum_{\alpha, \beta}((\epsilon-\mu) \delta_{\alpha \beta}- \vec{B} \vec{\sigma})d_{\alpha}^{\dag}d_{\beta} +V \sum_{\bm{k}, \alpha,i} e^{-i \vec k \vec R_i} \psi_{\alpha}^{\dag}(\bm{k})d_{\alpha} + {\rm h.c.}\label{ham},
\end{equation}
where 
\begin{equation}
h_{\alpha\beta}(\vec{k})=v_F(k_x\sigma_y-k_y'\sigma_x)_{\alpha\beta},
\end{equation}
is the (momentum shifted) Hamiltonian of the free Dirac fermions on the surface of the topological insulator.\\
From this we can identify the Green's function for the Dirac electrons
\begin{equation}
G^{\rm D}_{\alpha\beta}(\omega)=(\omega+\mu-\Sigma(\omega)-h(\vec{k}))^{-1}_{\alpha\beta},
\label{gdir}
\end{equation}
where the self-energy matrix $\Sigma(\omega)$ is due to scattering from the impurities which have a Green's function (T-matrix)
\begin{equation}
T(\omega)=V^2 g^{\rm imp}_{\alpha\beta}(\omega)=V^2(\omega+\mu-\epsilon-\Delta(\omega)+g\mu_{B} [\vec{B}\cdot\vec{\sigma}])^{-1}_{\alpha\beta},\label{gimp}
\end{equation}
where $\Delta(\omega)$ describes the hybridization of the impurity state with the continuum of Dirac electrons. $G^D$, $g^{\rm imp}$, $\Sigma$, and $\Delta$
are all $2 \times 2$ matrices. When no magnetization of the impurity is present (i.e. $B=0$), the self-energy $\Sigma(\omega)$ and hybridization $\Delta(\omega)$ are diagonal in spin-space.

To calculate the self-energy and hybridization function appearing in Eqs. \eqref{gdir} and \eqref{gimp} we take the first order of an $n^{\rm imp}$ expansion of these quantities. This corresponds to scattering events arising from only a single impurity. Hence, diagrammatically (see below), the irreducible contributions involve only impurity lines from a single scattering center and this approximation is known as the self-consistent T-matrix approximation (SCTMA) \cite{Mahan}. The SCTMA becomes mathematically exact in the limit that the ratio of the density of impurities $n^{\rm imp}$ to the density of electron states $\rho(\mu)$ becomes zero. This condition is not satisfied near the Dirac point where the density of states of the clean system is zero and so the SCTMA is not rigorously valid here. Despite this it has been shown that such an approximation accurately captures the qualitative physics of the a Dirac system coupled to impurities in the metallic regime far from the Dirac as well as in the impurity dominated regime close to the Dirac point \cite{Mirlin}.

Hence within the SCTMA the self-energy is given by
\begin{equation}
\begin{aligned}
&\Sigma_{\alpha\beta}(\omega)=n^{\rm imp}\vcenter{\hbox{
\includegraphics[scale=0.5]{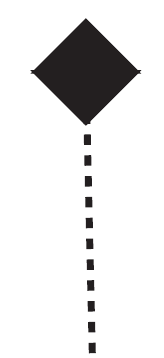}}}
=
n^{\rm imp} \left(\:\vcenter{\hbox{\includegraphics[scale=0.5]{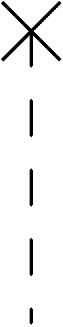}}}
+\vcenter{\hbox{\includegraphics[scale=0.5]{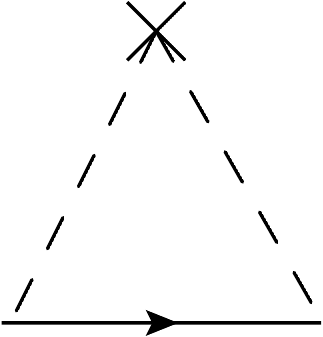}}}
+\vcenter{\hbox{\includegraphics[scale=0.5]{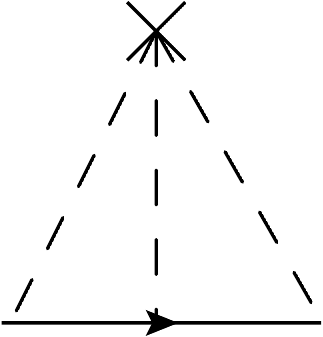}}}
+... \:\right)\\
&=n^{\rm imp} |V|^2 \langle g^{\rm imp}_{\alpha\beta}(\omega) \rangle_{\rm imp}=n_i |V|^2(\omega+\mu-\epsilon-|V|^2\int \frac{d{\bf k}}{(2\pi)^2} G({\bf k},\omega)+g\mu_{B} [\vec{B}\cdot\vec{\sigma}])^{-1}_{\alpha\beta},\label{self-T}
\end{aligned}
\end{equation}
where the average $\langle . \rangle_{\rm imp}$ is over the (random) positions of impurities and the hybridization matrix is given by
\begin{equation}
\Delta_{\alpha \beta}(\omega)= |V|^2 \int \frac{d{\bf k}}{(2\pi)^2} G^{\rm D}_{\alpha\beta}(\omega, \vec{k}). 
\end{equation}
The Dirac Green's function $G({\bf k},\omega)$ appearing in the hybridization function of $\langle g^{\rm imp}_{\alpha\beta}(\omega) \rangle_{\rm imp}$ includes $\Sigma(\omega)$ and so Eq.~\eqref{self-T} is a self-consistent equation for the self-energy. For non-zero fields Eq.~\eqref{self-T} is a $2\times2$ matrix equation.

\vskip 0.3cm
{\bf Zero-field self-energy:}
When $B=0$ the angular integral over the off-diagonal components of the Dirac Green's function cancels in the self-consistent equation Eq.~\eqref{self-T}, and the off-diagonal component of self-energy $\Sigma_{12}$ becomes zero. The remaining diagonal element $\Sigma_{11}=\Sigma_{22}$ is then given by \cite{Mirlin}
\begin{equation}
\begin{aligned}
\Sigma_{11}(\omega)&=n^{imp} |V|^2\left(\omega+\mu-\epsilon-|V|^2\int\limits_0^{\Lambda} \frac{d k}{2\pi} \frac{\omega+\mu-i\delta-\Sigma(\omega)}{(\omega+\mu-i\delta-\Sigma(\omega))^2-v_f ^2 k^2} \right) ^{-1}\\
&\approx n^{imp} |V|^2\left(\omega+\mu-\epsilon+\frac{|V|^2}{4 \pi v_f^2} (\omega+\mu-\Sigma(\omega))\ln\left(\frac{-\Lambda^2}{(\omega+\mu-\Sigma(\omega))^2}\right)\right)^{-1},
\end{aligned}\label{selfconana}
\end{equation}
where we have introduced a cut-off $\Lambda$ due to the logarithmic behavior of the integral. At zero frequency the equation has two distinct regimes: (i) In the metallic regime, where $|\mu|$ is large, both real and imaginary parts of self-energy are small, $\Sigma \sim 1/\mu$. (ii) An impurity dominated regime near the Dirac point, where the self-consistency becomes important. 

\begin{figure}
\begin{center}
\includegraphics[width=16cm]{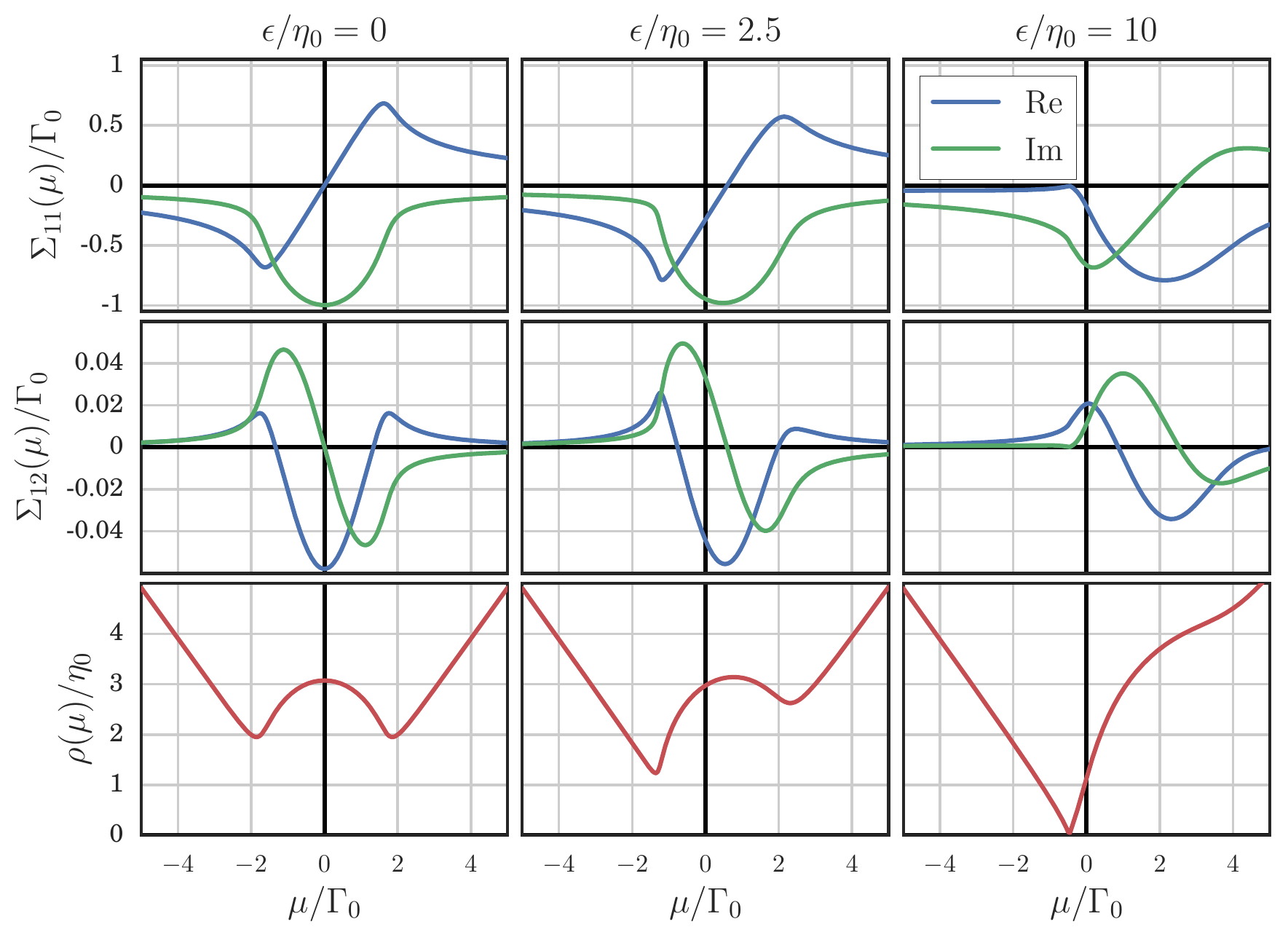}
\caption{Self energies and density of states as function of the chemical potential and anisotropy $\epsilon$.
Top row: The diagonal component of self-energy $\Sigma_{11}(\mu)$ for $\epsilon/\eta_0=0,\:2.5,\:10$ respectively. Middle row: The off-diagonal component of self-energy $\Sigma_{12}(\mu)\sim (\Sigma_{11}(\mu))^2$ for $B=0.25\eta_0$ and the same $\epsilon$ parameters. Bottom row: density of states. Impurity scattering dominates close to the Dirac point which leads to distinct features in all quantities. Parameters: $\Lambda=10$, $V=\sqrt{4 \pi}$.
\label{fig:self}} 
\end{center}
\end{figure}

Three examples of real and imaginary part of the self-energy and spectral functions are shown in Fig.~\ref{fig:self}. The impurity dominated regime is characterized by a large increase in the absolute value of the imaginary part of the self-energy; correspondingly, due to Kramers-Kr\"onig relation, there are two maxima in the real part of the self-energy. These peaks are associated with dips in the spectral functions.

For the particle-hole symmetric situation, $\epsilon=0$, Eq.~\eqref{selfconana} is purely imaginary at the center of the impurity dominated regime at $\mu=0$. The energy scale here, $\Sigma(\mu=0)=-i \Gamma_0$, defines the width of this impurity dominated regime. From Eq.~\eqref{selfconana} we see that $\Gamma_0$ is given by self-consistently solving
\begin{equation}
\Gamma_0=\sqrt{\frac{2\pi n^{\rm imp}v_f^2}{\ln\left(\frac{\Lambda}{\Gamma_0}\right)}}.
\end{equation}

For our discussion it is important to distinguish between weakly and strongly scattering impurities.
Within our model, strong impurity scattering is realized for small $\epsilon$, when scattering is approximately resonant. An inspection of the denominator in Eq.~\eqref{selfconana} reveals that strong, approximately resonant impurity scattering is realized for $|\epsilon| \lesssim |V|^2 \Gamma_0^2/(4 \pi v_f^2)=\eta_0$. In contrast, when $\epsilon \gg \eta_0 $, scattering from off-resonance impurities is weak.

The density of states is given by the trace of the Dirac Green's function at zero frequency,
\begin{equation}
\begin{aligned}
\rho(\mu)&=-\frac{1}{\pi} {\rm Im} \left\{ {\rm Tr} \int\frac{d^2 k}{(2\pi)^2}G(\mu,{\bf k}) \right\}={\rm Im} \left\{\frac{2 n^{\rm imp} |V|^2}{\pi \Sigma(\mu)}\right\}, \label{DOS}
\end{aligned}
\end{equation}
where the second line can be obtained by inserting the self-consistent equation Eq.~\eqref{selfconana} solved for the hybridization function. The density of states is also shown in Fig.~\ref{fig:self} for the same resonance values. From this we see that in the metallic regime the density of states is linear, as in the clean system. However close to the Dirac point the density of states is strongly affected by the presence of impurities with new states created between the bounds of the regime set by the energy scale $\Gamma_0$.

\vskip 0.3cm
{\bf Finite-field self-energy:}
At finite ${\vec B}$ (taking the $\|$-direction as the $x$-direction) the self-consistent T-matrix equation, Eq.~\eqref{self-T}, becomes a full matrix equation with non-zero off-diagonals of the self-energy matrix $\Sigma(\omega)$ and hybridization matrix $\Delta(\omega)$. This matrix equation is now 
\begin{equation}
\begin{aligned}
\Sigma(\omega)&=n_i |V|^2 \left(\begin{array}{cc}
\omega+\mu-\Delta_{11}(\omega)-\epsilon & g\mu_B B -\Delta_{12}(\omega)\\
g\mu_B B -\Delta_{12}(\omega) & \omega+\mu-\Delta_{11}(\omega)-\epsilon \\
\end{array}\right)^{-1}\\
&\approx \Sigma_0 (\omega)\mathbb{1} + g \mu_B B \frac{\Sigma_0(\omega)^2}{n_{\rm imp} |V|^2} \sigma_x,\label{selfB}
\end{aligned}
\end{equation}
where the last line is valid for small ${\vec B}$ and $\Sigma_0(\omega)$ is defined using the zero-field self-consistent equation
\begin{equation}
\Sigma_0(\omega)=\frac{n_i |V|^2}{\omega+\mu-\epsilon-\Delta_{11}(\omega)}.
\end{equation}
Examples of the real and imaginary part of $\Sigma_{12}(\omega)$ are shown in Fig.~\ref{fig:self}.

As discussed in the main text (see discussion below Eq. (2) there), our main experimental finding, the two-peak structure in the anisotropic magnetoresistance, can be traced back to the second line in Eq. \eqref{selfB}: ${\rm Im} \Sigma_{12}(\mu)$ is proportional to ${\rm Im}[\Sigma_{11}(\mu)^2]=2 {\rm Im}\Sigma_{11}{\rm Re}\Sigma_{11}$. Ultimately, this implies that the peaks in ${\rm Re}\Sigma_{11}$ lead to peaks in the gate-voltage dependence of the anisotropy, see main text.

Note, however, that the two peaks in ${\rm Im}\Sigma_{12}(\mu)$ and the related two-peak structure found in the conductivity (see main text) will vanish in a regime where all impurities are weakly scattering. In this Born limit (reached nominally in our model for $\epsilon \gg \eta_0$), where the impurity dominated regime is exponentially suppressed, $\Sigma_{12}(\mu)$ is proportional to 
$\rho(\mu)$ and no peaks will be visible.

\begin{flushleft} 
{\bf S6. Conductivity within SCTMA}
\end{flushleft} 

The DC conductivity is given by the Kubo formula \cite{Mahan}
\begin{equation}
\sigma^{\alpha\beta}(\mu)=\lim_{\Omega\rightarrow 0} \frac{1}{\Omega}\int \frac{d^2\vec{k}}{(2\pi)^2} \int_0^{\infty}dt\; e^{i\Omega t} \;Tr\langle [\hat{J}^{\alpha}(t),\hat{J}^{\beta}(0)]\rangle,
\end{equation}
where the current operator for spin-momentum locked surface states is given by 
$\vec{J}=e \frac{\partial H}{\partial {\bf k}}=e v_F\left(
\sigma_{y},
-\sigma_{x}
\right)$.
Ignoring for the moment vertex corrections (see below), the conductivity at $T=0$ is given by
\begin{equation}
\begin{aligned}
\sigma^{\alpha\beta}_0(\mu)&=\lim_{\Omega\rightarrow 0}{\rm Im} \left\{\frac{n_F(\omega+\Omega)-n_F(\omega)}{\Omega}\Pi_0(\omega+\Omega)\right\} \\
&=-{\rm Im} \left\{\int \frac{d \omega}{i\pi} \frac{\partial n_F(\omega)}{\partial \omega} \int \frac{d^2 {\bf k}}{(2\pi)^2} \text{Tr} \langle \hat{J}^\alpha G({\bf k}, \omega) \hat{J}^\beta G^{\dagger}({\bf k},\omega)\rangle \right\}
\\&=-{\rm Im} \left\{\frac{e^2 v_F^2}{i\pi} \int \frac{d^2 {\bf k}}{(2\pi)^2} \text{Tr} \langle \sigma^\alpha G({\bf k}, \mu) \sigma^\beta G^{\dagger}({\bf k},\mu)\rangle \right\}.
\end{aligned} \label{novertex}
\end{equation}
Additional terms of the form $ \langle \hat{J}^\alpha G \hat{J}^\beta G \rangle$ and $ \langle \hat{J}^\alpha G^{\dagger} \hat{J}^\beta G^{\dagger}\rangle$ will be present but are equal for $\sigma^{\|}$ and $\sigma^{\perp}$ and so do not contribute to the anisotropy in MR. This is because ${\rm Im}\Sigma_{12}$ always has the same sign for these terms and so can be completely eliminated by a shift in $k_y$. In all three types of terms ${\rm Re}\Sigma_{12}$ can also be eliminated in a similar manner and so does not contribute to the AMR. The resulting difference in conductivity can be interpreted in terms of the ratio between spin-flip and non-spin-flip scattering (see main text). 

For the particle-hole symmetric system (i.e. $\epsilon=0$), the conductivity at the Dirac point within the SCTMA is $e^2/2\pi^2$. This is a quarter the value found in graphene (which has an additional two valley and spin degrees of freedom) within the same approximation \cite{Mirlin}. 

\vskip 0.3cm
{\bf Vertex corrections:}
The vertex corrections to the conductivity of graphene vanish for short-ranged impurities \cite{Mirlin}. The locking of spin and momentum for surface states of topological insulators implies that impurity scattering is always angular dependent. This implies that vertex corrections do not vanish in this case and have to be taken into account within the self-consistent T-matrix approximation. Within the SCTMA framework, vertex corrections are obtained from a sum of ladder diagrams \cite{Mahan,Mirlin}, 
\begin{equation}
\vcenter{\hbox{\includegraphics[scale=0.1]{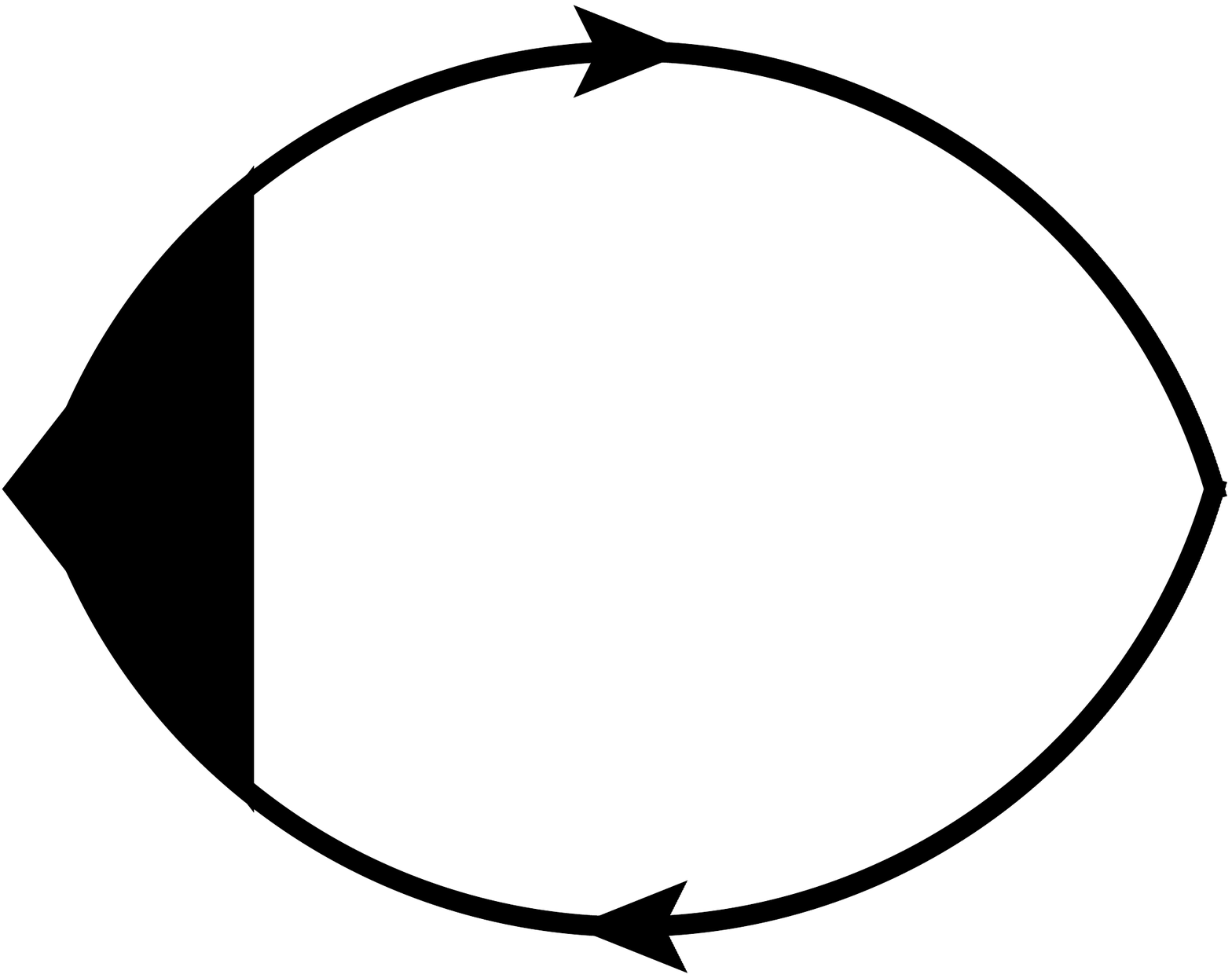}}}=
\:\vcenter{\hbox{\includegraphics[scale=0.1]{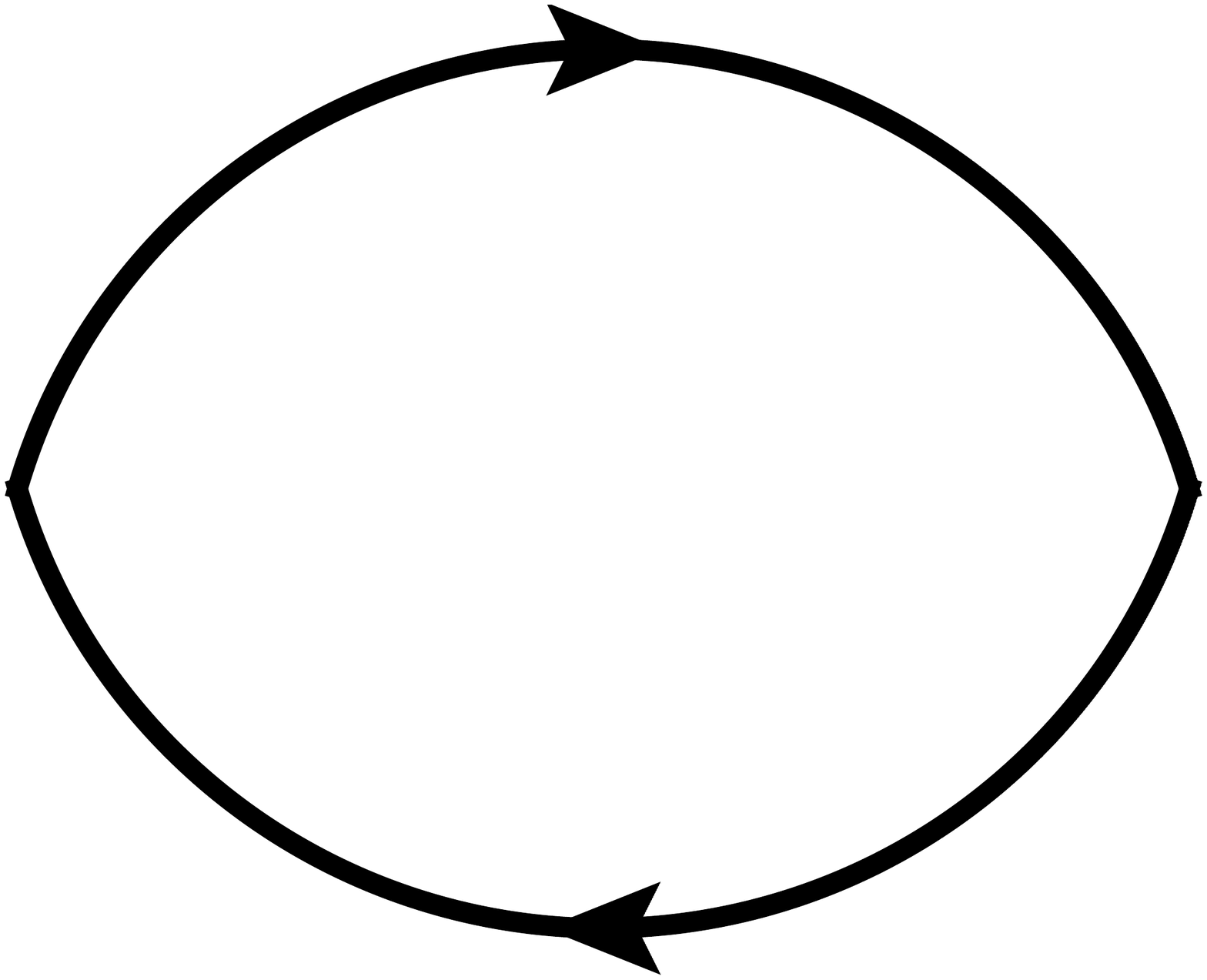}}}
+\vcenter{\hbox{\includegraphics[scale=0.1]{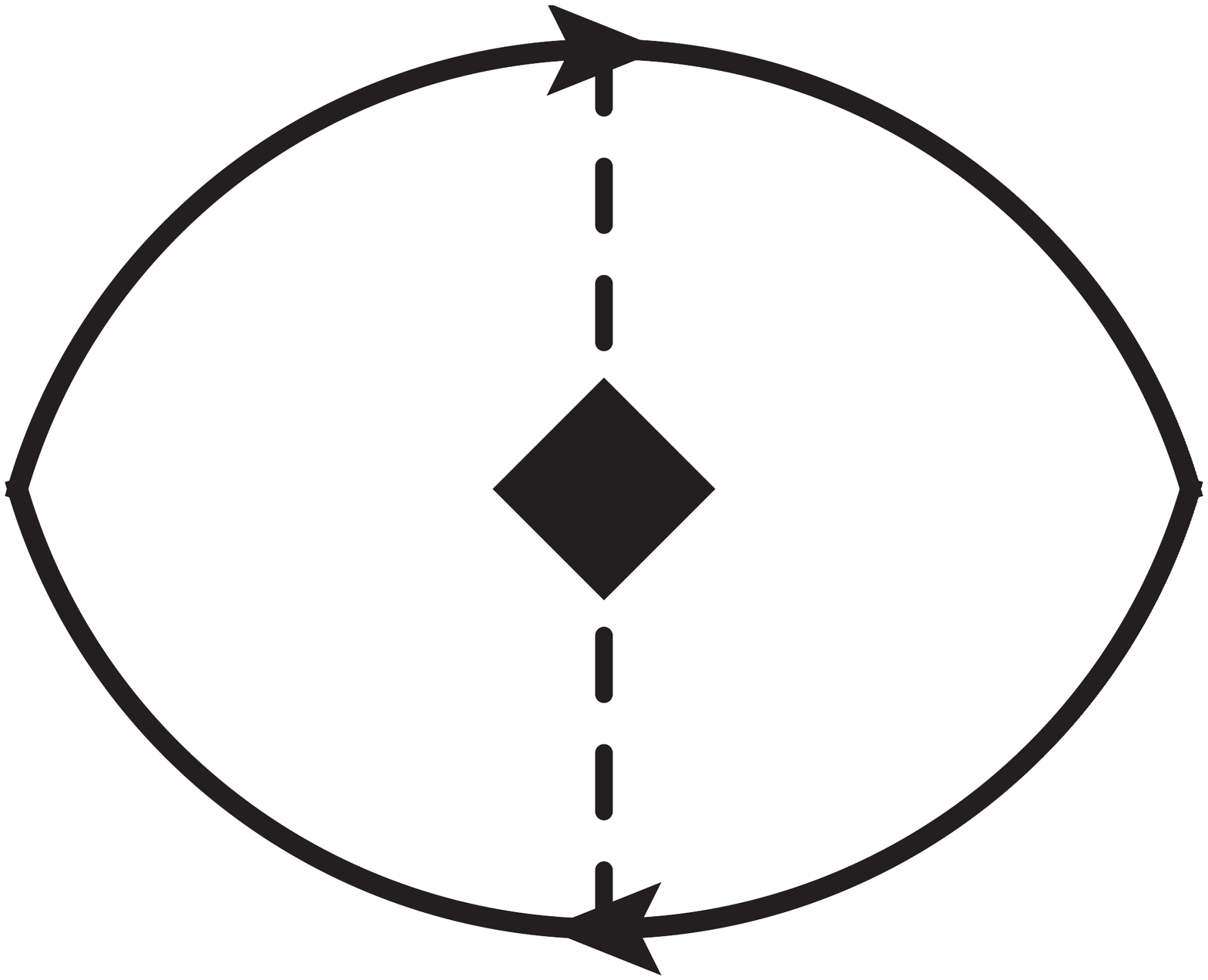}}}
+\vcenter{\hbox{\includegraphics[scale=0.1]{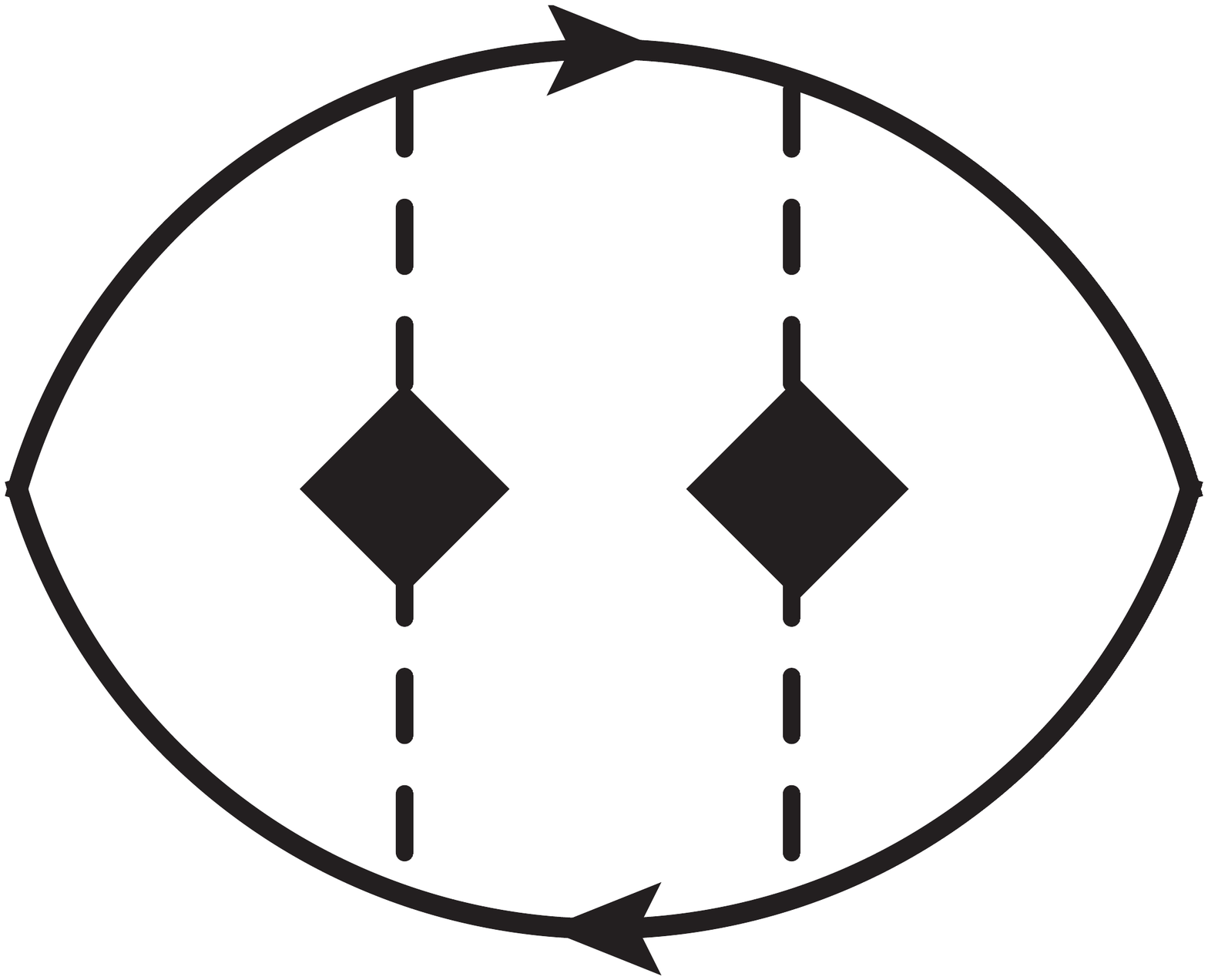}}}
+... \:\label{ladders}
\end{equation}
Fortunately, it is not necessary to solve an integral equation to resum the vertex corrections within our model.
Instead, one can use the following trick: All $\vec k$ summations can directly be done by
defining the $4\times4$ matrices 
\begin{equation}
\mathcal{M}(\omega,\Omega)=\int \frac{d^2 {\bf k}}{(2\pi)^2} G({\bf k}, \omega+\Omega) \otimes G^{\dagger}({\bf k}, \omega) 
\end{equation}
and
\begin{equation}
\mathcal{T}(\omega,\Omega)=T( \omega+\Omega) \otimes T(\omega), \label{TT}
\end{equation}
where $\otimes$ is the Kronecker product. Within this space the Pauli-matrices in the current vertex map to column vectors and Eq.~\eqref{ladders} can be written in terms of a geometric series of $4\times4$ matrices
\begin{equation}
\begin{aligned}
\sigma^{\alpha\beta}(\mu)&=\lim_{\Omega\rightarrow 0}{\rm Re}\bigg\{e^2 v_F^2 \int \frac{d \omega}{\pi} \frac{\Delta n_F(\omega)}{\Omega} \sigma^{\alpha}.(\mathcal{M}+n^{\rm imp}\mathcal{M}\mathcal{T}\mathcal{M}+...).\sigma^{\beta}\bigg\}\\
&=\lim_{\Omega\rightarrow 0}{\rm Re}\bigg\{e^2 v_F^2 \int \frac{d \omega}{\pi} \frac{\Delta n_F(\omega)}{\Omega} \sigma^{\alpha}.(\mathcal{M}.(\mathbb{1}_4-n^{\rm imp}\mathcal{T}\mathcal{M})^{-1}).\sigma^{\beta}\bigg\}.
\end{aligned}\label{vertex}
\end{equation}
To be precise, the formulas given above are only complete when one calculates the anisotropy of the resistivity, $\sigma_\perp-\sigma_\|$. Otherwise one has also to include extra isotropic contributions arising from contributions where either $G^\dagger$ is replaced by $G$ or $G$ by $G^\dagger$, see discussion below Eq.~\eqref{novertex}.

We would like to emphasize that the limit $\Omega\rightarrow 0$ has to be taken with some care (i.e. only at the very end of the calculation) in this Dirac system. This is related to the fact that for finite magnetization $\langle \hat{J}^x \rangle(\mu)\neq 0$ even for vanishing electric field.

\begin{figure}
\begin{center}
\includegraphics[width=16cm]{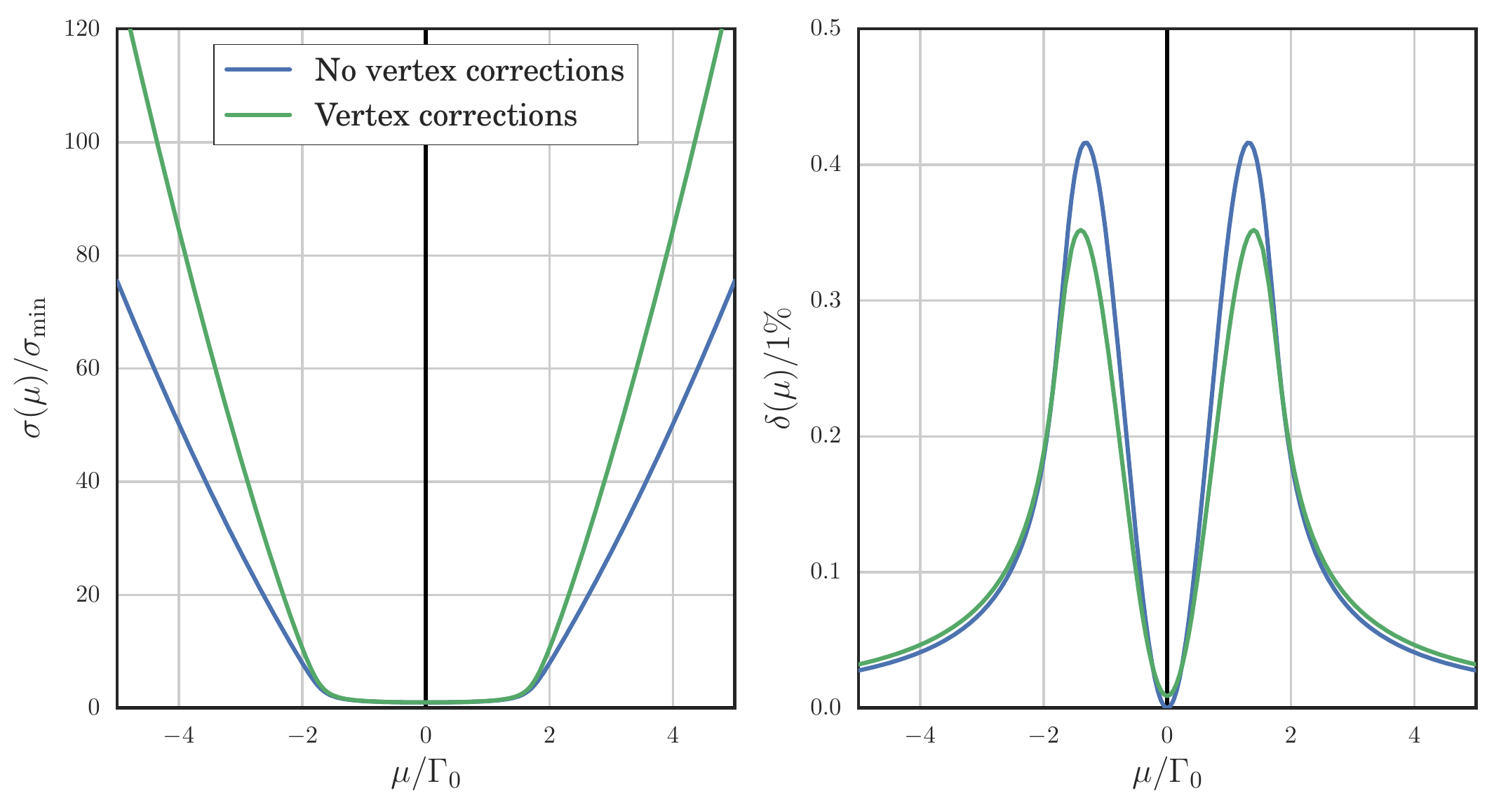}
\caption{
The conductivity (left) and dimensionless resistivity anisotropy $\delta(\mu)=(\sigma_\perp-\sigma_\|)/\sigma_\|$ (right) with and without vertex corrections for the particle hole symmetric system $\epsilon=0$. Vertex corrections approximately double the conductivity in the metallic regime but have only a small effect near the peaks of $\delta(\mu)$. \label{fig:vertex}} 
\end{center}
\end{figure}

As can be seen from Fig.~\ref{fig:vertex}, the vertex corrections approximately double the conductivity in the metallic regime far from the Dirac point, but for the ratio $\delta(\mu)=(\sigma_\perp-\sigma_\|)/\sigma_\|$ the vertex corrections cause only a small reduction in the vicinity of the peaks.

\vskip 0.3cm
{\bf Averaging over impurity parameters:}
In reality the topological insulator surface may contain different types of impurities described, for example, by a distribution of parameters $V$ and $\epsilon$. To check the robustness of our description of the experiment, we therefore show in the following that such distributions do not affect our conclusions. 

\begin{figure}
\begin{center}
\includegraphics[width=16cm]{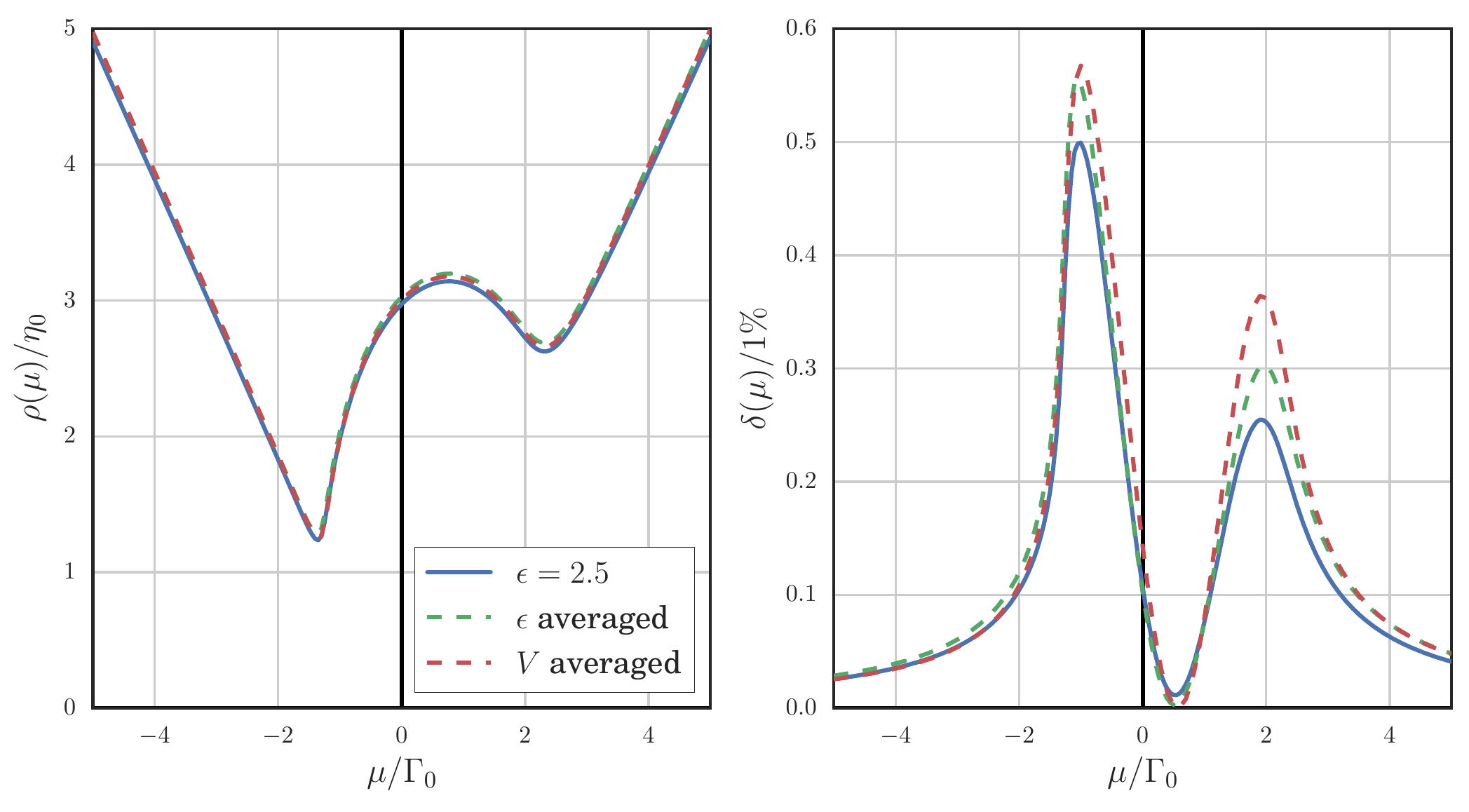}
\caption{
The density of states (left) and resistivity anisotropy $\delta(\mu)$ (right) as a function of chemical potential for a single type of impurity with $\epsilon=2.5 \eta_0$, $V=\sqrt{4 \pi}$ (blue line), a Gaussian distribution of $\epsilon$ values with $\bar \epsilon=2.5 \eta_0$ and width $\Delta_\epsilon=\eta_0$ (green, dashed), and a Gaussian distribution of $V$ values with $\bar V=\sqrt{4 \pi}$ and width $\Delta_V=\bar V/8$ (red, dashed). We see that averaging has a negligible effect on the density of states and on the position of the peaks in $\delta(\mu)$. The peaks are, however, slightly broadened and the size of the peaks increases as impurities with smaller $\epsilon$ and smaller $V$ get magnetized more strongly, leading to enhanced spin-flip scattering.\label{fig:avg}} 
\end{center}
\end{figure}

Within the SCTMA, the average over parameters can directly be implemented by averaging over $\epsilon$ and/or $V$ in Eq.~\eqref{self-T}. 
To perform this averaging we assume that the distribution of parameters is described by a Gaussion distribution $n(\epsilon,V)$ 
with averages $\bar \epsilon$ and $\bar V$, widths $\Delta_\epsilon$ and $\Delta_V$, and $\int n(\epsilon,V) \, d\epsilon \,dV= n^{\rm imp}$.
To calculate vertex corrections, Eq.~\eqref{TT} has to be replaced by
\begin{equation}
\mathcal{T}(\omega,\Omega)=\int d\epsilon\, dV\, \frac{n(\epsilon,V)}{n^{\rm imp}} T(\omega+\Omega) \otimes T(\omega).
\end{equation}
Figure~\ref{fig:avg} shows the results of both $V$ and $\epsilon$ averaging. We see that the distribution has no effect on the position of the peaks in $\delta(\mu)$, which are only slightly broadened. The height of the peaks in increased due to the enhanced magnetic scattering arising from impurities with smaller $\epsilon$ and $V$. Most importantly, averaging over impurity distributions does not affect our interpretation of the anisotropic magnetoresistance put forward in the main text.

\end{document}